\documentclass{aa}  

\usepackage{graphicx}
\usepackage{siunitx}
\usepackage{float}
\usepackage{xcolor}
\usepackage{amsmath}
\usepackage{subcaption}
\usepackage{txfonts}
\begin{document} 

    \title{Modelling Solar Orbiter Dust Detection Rates \\ in Inner Heliosphere as a Poisson Process }



   \author{S. Ko\v{c}i\v{s}\v{c}\'{a}k
          \inst{1}\fnmsep\thanks{samuel.kociscak@uit.no}
          \and
          A. Kvammen
          \inst{1}
          \and
          I. Mann
          \inst{1}
          \and
          S. H. S{\o}rbye
          \inst{2}
          \and
          A. Theodorsen
          \inst{1}
          \and
          A. Zaslavsky
          \inst{3}
          }

   \institute{
   Department of Physics and Technology, UiT The Arctic University of Norway, 9037, Troms{\o}, Norway
         \and
    Department of Mathematics and Statistics, UiT The Arctic University of Norway, 9037, Troms{\o}, Norway
        \and
    LESIA, Observatoire de Paris, Universit\'{e} PSL, CNRS, Sorbonne Universit\'{e}, Universit\'{e} de Paris, Paris, France
            }

   \date{Received Month DD, YYYY; accepted Month DD, YYYY}

 
  \abstract
   {Solar Orbiter provides dust detection capability in inner heliosphere, but estimating physical properties of detected dust from the collected data is far from straightforward.}
   {First, a physical model for dust collection considering a Poisson process is formulated. Second, it is shown that dust on hyperbolic orbits is responsible for the majority of dust detections with Solar Orbiter's Radio and Plasma Waves (SolO/RPW). Third, the model for dust counts is fitted to SolO/RPW data and parameters of the dust are inferred, namely: radial velocity, hyperbolic meteoroids predominance, and solar radiation pressure to gravity ratio as well as uncertainties of these.}
   {Non-parametric model fitting is used to get the difference between inbound and outbound detection rate and dust radial velocity is thus estimated. 
   A hierarchical Bayesian model is formulated and applied to available SolO/RPW data. The model uses the methodology of Integrated Nested Laplace Approximation, estimating parameters of dust and their uncertainties.}
   {SolO/RPW dust observations can be modelled as a Poisson process in a Bayesian framework and observations up to this date are consistent with the hyperbolic dust model with an additional background component. Analysis suggests a radial velocity of the hyperbolic component around $\SI[separate-uncertainty]{63 \pm 7}{km/s}$ with the predominance of hyperbolic dust about $\SI[separate-uncertainty]{78 \pm 4}{\%}$. The results are consistent with hyperbolic meteoroids originating between $\SI{0.02}{AU}$ and $\SI{0.1}{AU}$ and showing substantial deceleration, which implies effective solar radiation pressure to gravity ratio $\gtrsim 0.5$. The flux of hyperbolic component at $\SI{1}{AU}$ is found to be $\SI[separate-uncertainty]{1.1(2) e-4}{m^{-2}s^{-1}}$ and the flux of background component at $\SI{1}{AU}$ is found to be $\SI[separate-uncertainty]{5.4(15) e-5}{m^{-2}s^{-1}}$.}
   {}

   \keywords{   interplanetary medium --
                interplanetary dust --
                Solar Orbiter --
                methods: statistical -- 
                Bayesian inference
               }

   \maketitle
%
\section{Introduction}

\subsection{Hyperbolic dust and $\beta$-meteoroids}

Among dust detected with in-situ measurements within $\SI{1}{AU}$, particles on unbound hyperbolic trajectories originating in the relative vicinity of the Sun play a major role, as was already shown in case of measurements of Solar Orbiter \citep{zaslavsky2021first}. Most of these hyperbolic particles of sub-micron size are believed to be so-called $\beta$-meteoroids, generated by a high radiation pressure to gravity ratio, denoted $\beta$:  

\begin{equation}
    \qquad \qquad \qquad \qquad \beta = \frac{F_{radiation}}{F_{gravity}}.
\end{equation}

It is clear that in the region of dust sizes $s \gg \lambda$, where $s$ denotes the dimension of a dust particle and $\lambda$ denotes the wavelength of incident light, $F_{radiation}$ depends on dust particle's cross section, while $F_{gravity}$ depends on dust grain's volume. Hence, the smaller the particle, the higher the $\beta$ value. Maximum of $\beta$ is therefore reached when $s \approx \lambda$ and usually $\beta_{max} \approx 1$. Notably, both $F_{radiation}$ and $F_{gravity}$ depend on the inverse square of heliocentric distance, hence $\beta$ remains constant for a given particle throughout its trajectory. Note that $F_{radiation}$ and $F_{gravity}$ are the predominant forces for the $\beta$-meteoroids, as electromagnetic forces become relevant for dust grains of size $s < \SI{100}{nm}$ \citep{mann2014dust}. Dust of size $s < \SI{100}{nm}$ can also be on an unbound trajectory due to electromagnetic forces \citep{czechowski2021dynamics,mann2021dust}.

For $\beta = 1$, the grain neither accelerates nor decelerates due to Solar influence. For $\beta = 0.5$, effective Solar attraction is reduced to $1/2$, which means that a sudden change in $\beta$ from $0$ to $0.5$ will cause an originally circular orbit to become an unbound, parabolic orbit. Particles with $\beta \approx 1$ are created mostly in collisions of larger dust \citep{dohnanyi1972interplanetary,zook1975source,grun1985collisional}. Larger particles have very low $\beta \ll 1$ and are therefore originally on Keplerian orbits (referred to as initial orbits hereafter), hence $\beta = 0.5$ could be considered the minimal value needed for dust to become unbound.

The population of bound ($\beta \ll 0.5$) dust particles inside $\SI{1}{AU}$ is notably responsible for visual observations of zodiacal light. Their spatial density has been observed to depend on heliocentric distance approximately as $\sim r^{-1.3}$ \citep{leinert1981zodiacal}, which holds well down to $\SI{20}{R_\odot}$, or $\SI{0.1}{AU}$. Inward of that distance, they show shallower dependence, suggesting a maximum in density somewhere inward of $\SI{0.05}{AU}$, or $\SI{10}{R_\odot}$ \citep{stenborg2021psp}. Regions with high density of bound dust is very likely the region of origin of $\beta$-meteoroids, as the collision rate of bound dust depends on the square of its spatial density \citep{mann2005dust}. 

As $\beta$-meteoroids likely make up most of the submicron hyperbolic dust, the two terms are almost interchangeable for the purpose of the present discussion. The term \textit{$\beta$-meteoroids} is used where radiation pressure ejection is important and the term \textit{hyperbolic dust} is used where only trajectories of the grains are relevant.

\subsection{Impact ionization}

Detection of bound dust particles is usually done remotely, both historically \citep{van1947zodiacal,leinert1981zodiacal} and to current date \citep{howard2019near,stenborg2021psp} taking advantage of light scattering properties of these particles. Detection of sub-micron particles is done mostly in-situ (at encounter with a particle), due to their insignificant light scattering properties and low spatial density, often taking advantage of so-called impact ionization effect \citep{friichtenicht1962two,alexander1968zodiacal}. Impact ionization dust detection is a passive data gathering process carried out by either a specialized instrument \citep{dietzel1973heos,srama2004cassini} or often as a byproduct of electric \citep{gurnett1997micron,meyer1986voyager,kurth2006cassini,wang2006characteristics,zaslavsky2012interplanetary,vaverka2018comparison,malaspina2020situ,mozer2020time,zaslavsky2021first,nouzak2021detection} or magnetic \citep{malaspina2022clouds,gasque2022magnetic} measurements. Due to high energy density present at the impact site, free charge is generated upon hypervelocity dust impact. The charge generated is partially picked up by the spacecraft body and/or antennas, which results in specific signatures in fast electric measurements \citep{zaslavsky2015floating, meyer2017frequency, vaverka2017potential, mann2019dust, shen2021electrostatic, babic2022analytical}. The amount of generated charge $Q$ has been empirically found to approximately follow the equation 
\begin{equation}
    Q = A m^{\gamma} v^{\alpha} \label{eq:charge_generation},
\end{equation}
where in the range of impact velocities $20 < v < 50$ achieved in laboratory \citep{friichtenicht1962two,dietzel1972micrometeoroid,shu20123}, $\gamma \approx 1$ and $3 \lesssim \alpha \lesssim 5$. All three parameters $A$, $\gamma$, and $\alpha$ are dependent on both the material of the grain and the target \citep{grun1984impact,grun2007dust,collette2014micrometeoroid}.  

\subsection{Dust mass distribution observations}

For many decades now, it has been standard to express cumulative mass distribution of dust near $\SI{1}{AU}$ in terms of power-law distribution over about $20$ decades of masses, from nanodust to comets and asteroids and above. Clearly, the distribution is an approximation and the distribution is described with different exponent in different intervals. However, it is often the case that a single experiment is sensitive over several orders of magnitude and finds that the mass distribution (number of particles with mass at least $m$) follows a power-law
\begin{equation}
    F(m) = F(m_0) \left( \frac{m}{m_0} \right)^{-\delta}
    \label{eq:mass_distribution}
\end{equation}
over the observed range. For example, the work of \cite{whipple196756} reported $\delta \approx 1.34$ for the mass range $10^{-8} - 10^{-1} \si{kg}$ and $\delta \approx 0.51$ for the mass range $10^{-13} - 10^{-8} \si{kg}$. Compiling previous estimates and relying on the stationarity of dust cloud, \cite{dohnanyi1970origin} reported $\delta \approx 7/6$ for sporadic meteoroids of masses from macroscopic down to $10^{-11} \si{kg}$ and $\delta \approx 1/2$ between $10^{-14} \si{kg}$ and $10^{-11} \si{kg}$. \cite{grun1985collisional} suggested $\delta \approx 0.8$ in the range $10^{-21} - 10^{-17} \si{kg}$, that is $\beta$-meteoroids and smaller. Recently, \cite{zaslavsky2021first} inferred $\delta \approx 0.34$ for Solar Orbiter's Radio and Plasma Waves (SolO/RPW for short) dust detections of dust of $m \gtrsim 10^{-17} \si{kg}$. It is not clear, whether eq. \eqref{eq:mass_distribution} represents a good approximation for $\beta$-meteoroids. 

\subsection{Poisson point process} \label{ch:poisson}

A Poisson point process is a stochastic process defined by the following properties:

\begin{enumerate}
    \item Poisson distribution of counts within arbitrarily chosen bounded region (for example a temporal interval), 
    \item Statistical independence of counts within disjoint regions (temporal intervals), and
    \item No two events can happen at the exact same location (time).
\end{enumerate}

In case of dust detection in the solar system, we assume the third condition met, as the detection rate is much lower than the detection process duration. The first two conditions demand that a detection of a dust particle does not influence the probability of detection of any other particle in any other time point, for example particles do not interact, do not come in swarms, and their reservoir cannot be depleted. All these can be assumed, as particles are likely formed far away from the spacecraft, are sparsely distributed, and their trajectories are uncorrelated. In that case, a Poisson process is the simplest conceivable model and it is natural to consider dust counts as an inhomogeneous Poisson point process, that is a Poisson process with non-constant rate. This means that the rate depends on other parameters, in our case distance from the Sun and spacecraft velocity. In fact, the observed number of detections within a naturally considered temporal interval, for example an hour or a day, is usually a low number. This implies a considerable probability of $0$ detections, which makes the random variable of detections per temporal interval a poor fit to often considered normal distribution, which allows for negative numbers. Hence, Poisson distribution of counts should be considered.

\subsection{Bayesian inference}

Inference of variable detection rate could be done by least-squares fitting of a model onto a time-series of detections per unit time, as is often done. A least squares fit produces the maximum likelihood estimate when the error of the data (residuals) are normally distributed. However, detections per unit time have a Poisson distribution, as discussed in sec. \ref{ch:poisson}. It is possible to obtain a maximum likelihood estimate with more careful analysis, but uncertainty is not directly accessible and must be estimated by other means (for example bootstrap). Adapting a procedure designed specifically to fit a Poisson process to Poisson observations grants the resolution needed to fit a complicated model precisely. Moreover, given we meet model assumptions, we can make more meaningful error estimates and potentially compare competing models in a meaningful way.

In the present work we take advantage of Bayesian inference, which is a general procedure that works with models for observations with unknown parameters and meets both of the above-mentioned criteria: it handles the Poisson distribution and provides an uncertainty estimate. In this approach, unknown parameters are regarded as random variables coming from an unknown distribution, about which some prior information is available (in the form of a prior belief, or prior distribution). The procedure infers the posterior (improved) distribution of unknown parameters based on the prior distribution and observed data. This distribution automatically carries information about uncertainty. 

\subsection{Integrated Nested Laplace Approximation}

Integrated nested Laplace approximation (\cite{rue2009approximate, rue2017bayesian}, INLA for short) implements approximate Bayesian inference for a wide class of three-stage hierarchical models. This class of models contains multilevel (nested) models, spatio-temporal models, survival models, and others \citep{gomez2020bayesian}. A decisive advantage of INLA above other Bayesian methods (for example sampling-based methods) is its computational efficiency allowing for fitting more complicated models to more observations within available time, making it the method of choice for the course of this work. The inference is carried out using the R-INLA package \citep{martins2013bayesian, rue2017bayesian} for the Bayesian inference. 

\subsection{Paper structure}

In section \ref{ch:data} we briefly introduce the Solar Orbiter mission, its dust measurement results and the data product that we used throughout the work. Section \ref{ch:velocity} is a discussion and analysis of observed hyperbolic dust velocity. 
The fitting of dust detection rate using INLA is presented in section \ref{ch:inla} and we conclude our findings in section \ref{ch:conclusion}. Finally, an outlook for Solar Orbiter and other missions is briefly discussed in section \ref{ch:outlook}.

\section{Solar Orbiter's dust observations and data products} \label{ch:data}

Solar Orbiter (SolO) is a spacecraft that orbits the Sun on an elliptical trajectory. SolO underwent several gravity assists and its orbital parameters have therefore changed several times since its launch in early 2020. As of summer 2022, SolO has low inclination, effectively making measurements in the ecliptic plane. Its aphelion is close to $\SI{1}{AU}$ and perihelion $\SI{0.3}{AU}$, but for the majority of its mission so far, its perihelion was close to $\SI{0.5}{AU}$. 

Radio and Plasma Waves (RPW) is an experiment onboard SolO designed to measure both the  electric and the magnetic field in three components in a wide frequency band, from near-DC to $\SI{16}{MHz}$ in case of the electric field \citep{maksimovic2020solar}. The measurements of electric fields, crucially for the present work, allow for detection of cosmic dust impacts, which is one of the auxiliary scientific objectives of SolO. A part of the data analyzed here was accessed at \cite{rwp_db}, specifically TDS waveform electrical data (Level 2). 

In their recent work, \cite{zaslavsky2021first} described properties of SolO/RPW as a dust detector. It has the capacity $C \approx \SI{250}{pF}$, sensitivity to pulses $V \gtrsim \SI{5}{mV}$, a collection area of $S_{col} \approx \SI{8}{m^2}$, and a duty cycle of $D \approx \SI{6.2}{\%}$. They showed that SolO/RPW is indeed capable of dust detections and that these could be modelled as hyperbolic dust. The authors mostly discuss $\beta$-meteoroids, as they are likely the observed population, but in principle the model fits to any hyperbolic dust population. The model for the detection rate $R$ presented in aforementioned work,

\vspace{-30pt}

\begin{subequations}
\begin{align} 
\begin{split} \label{eq:z2_model}
    R =& \ F_{1AU} S_{col} \left( \frac{r}{\SI{1}{AU}} \right)^{-2} \frac{v_{impact}}{v_{dust}} \left( \frac{v_{impact}}{v_{impact}(\SI{1}{AU})} \right)^{\alpha \delta},
\end{split}\\
\begin{split}
    \qquad v_{impact} =& \ | \vec{v_{dust}} - \vec{v_{SolO}} |, \\
    =& \ \sqrt{\left(v_{dust}^{radial} - v_{SolO}^{radial}\right)^2 + \left(v_{dust}^{azimuthal} - v_{SolO}^{azimuthal}\right)^2}
\end{split}
\end{align}
\end{subequations}
has three parameters: $\alpha \delta$, $F_{1AU} S_{col}$, and $v_{dust}$. Note that, both $\alpha \delta$ and $F_{1AU} S_{col}$ are products of two quantities. All three parameters could have a spatio-temporal dependence. 

The model shows a good fit to the data with the parameters taken constant: outward radial heliocentric speed of $v_{dust} \approx \SI{50}{km/s}$ and exponent $\alpha \delta \approx 1.3$. The value for $v_\beta$ was inferred by relating the difference in detection rate in the inbound and the outbound leg of an orbit \citep{zaslavsky2021first} and plugged into equation \eqref{eq:z2_model}. A value of $\delta \approx 0.34$, which is a dimensionless parameter in mass distribution of detected dust grains (see eq. \eqref{eq:mass_distribution}), was inferred from the distribution of impact pulse amplitudes. The value of $\alpha$, which stands for power of velocity in charge-yield equation \eqref{eq:charge_generation} is deduced from the knowledge of $\alpha \delta$ and $\delta$. $\alpha$ is often measured in laboratory setup and its inferred value is compatible with ground-based measurements \citep{collette2014micrometeoroid}. The arameter $F_{1AU}$ stands for the flux at $\SI{1}{AU}$, $F_{1AU} \approx \SI{8e-5}{m^-2 s^-1}$, see \cite{zaslavsky2021first} for details. 

In addition to L2 SolO/RPW data, this work makes use of the data product provided by \cite{kvammen2022convolutional}, which is a result of a convolutional neural network classified time domain sampled data. It builds on a supervised classification algorithm trained using a randomly chosen sub-sample of manually labelled data. Its main advantage over (in itself time consuming) visual inspection of all data automatically classified as dust is that it is fully automatic and reasonably time consuming. Therefore it allows for not only type 1 error correction (detection confirmation), but also for type 2 error correction implying search for dust in the vast data that was not classified as dust by an on-board algorithm a priori. Although no supervised classifier could get rid of human bias and error completely, this data provides the most reliable SolO dust detection data available to this date, as was shown in \cite{kvammen2022convolutional}. The data set consists of $4606$ dust detections aquired over approx. $669$ hours within $457$ days. We refer to this data as TDS/TSWF-E/CNN and it is publicly accessible, see \cite{kvammen2022convolutional}.

\section{Impact rates and velocities of hyperbolic dust} \label{ch:velocity}

\subsection{Single-particle velocities}

\begin{figure}[h]
\centering
\includegraphics{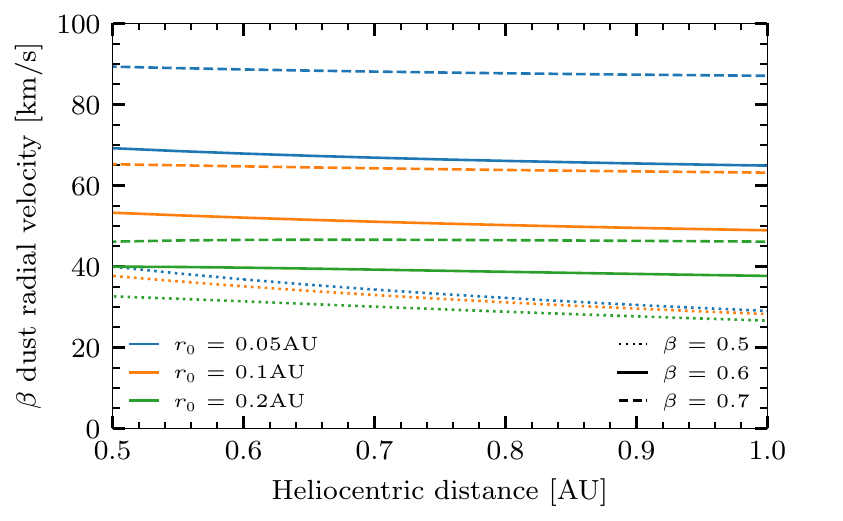}
\caption{Radial velocity profiles of $\beta$-meteoroids released by a sudden parameter change (for example due to a collision) from an initially circular orbit. A selection of $\beta$ values and origins ($r_0$) is shown.}
\label{fig:plt:beta_rad_velocity}
\end{figure}

$\beta$-meteoroids are moving mostly radially outward from their region of origin, which is located well within $\SI{0.5}{AU}$. Figs. \ref{fig:plt:beta_rad_velocity}, \ref{fig:plt:beta_azimuthal_velocity} display possible single-particle velocity profiles, see appendix. \ref{app:single_particle_valocities} for underlying equations. As $\beta < 0.5$ leads to finite aphelion, while $\beta \approx 1$ requires a rather specific set of parameters, values of $\beta \gtrsim 0.5$ are shown. Note that this choice is inconsequential and for illustration purposes only, as we do not presuppose a $\beta$ value in further analysis. In fact, we do not presuppose that the observed population is $\beta$-meteoroids, though that is likely the case. The effective initial orbit of $\beta$ grain's parent body must lie outside of near-solar dust-free zone, but in the region with high bound dust concentration, which restrains the $r_0$ values shown. As shown in Fig. \ref{fig:plt:beta_rad_velocity}, radial $\beta$-meteoroid velocities expected between $\SI{0.5}{AU}$ and $\SI{1}{AU}$ are between $\SI{30}{km/s}$ and $\SI{90}{km/s}$ for the given combinations of parameters and nearly independent of heliocentric distance (nearly constant). Solar gravity and radiation pressure forces are central forces, therefore the $\beta$ value does not influence azimuthal velocity as a function of heliocentric distance, which is governed by angular momentum conservation and initial orbit only. Azimuthal velocities of $\beta$-meteoroids of chosen parameters between $\SI{0.5}{AU}$ and $\SI{1}{AU}$ are therefore between $\SI{7}{km/s}$ and $\SI{30}{km/s}$ and decreasing $\propto r^{-1}$, as shown in Fig. \ref{fig:plt:beta_azimuthal_velocity}.

If dust detections on SolO/RPW correspond to hyperbolic dust, a difference in detection rate $R_{in}$ vs. $R_{out}$ due to spacecraft radial velocity should be present, as indeed is the case. It was shown by \cite{zaslavsky2021first} that this approach allows for order of magnitude estimation of the radial component of dust velocity $v_{dust;rad} \approx \SI{50}{km/s}$, which is in line with expectations. In the present work, we extend the approach to estimate continuous heliocentric distance dependent dust radial velocities, using the data product of \cite{kvammen2022convolutional} and taking into account more unknown variables that influence our estimates. 

\begin{figure}[h]
\centering
\includegraphics{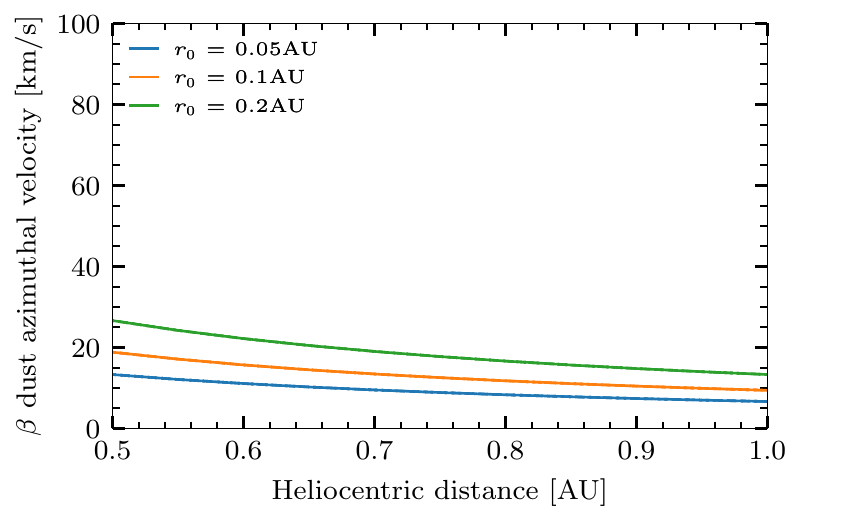}
\caption{Azimuthal velocity profiles of $\beta$-meteoroids released by a sudden parameter change (for example due to a collision) from an initially circular orbit. A selection of origins ($r_0$) is shown, the $\beta$ value is not relevant.}
\label{fig:plt:beta_azimuthal_velocity}
\end{figure}

\subsection{Velocity estimation} \label{ch:velocity_nonparametric}

A first estimate of $v_{dust;rad}$ is obtained if a model for dust collection rate with linear dependence on relative SolO and dust velocity $v_{impact}$ is assumed ($R \propto v_{impact}$). This corresponds to linear dependence on volume of space scanned per unit of time only:
\begin{equation}
    v_{dust;rad} \approx \frac{R_{in}+R_{out}}{R_{in}-R_{out}} | v_{sc;rad} |,
    \label{eq:velocity_first_estimate}
\end{equation}
where $| v_{sc;rad} |$ is the absolute value of the spacecraft's radial velocity at a given heliocentric distance. $R_{in}$ and $R_{out}$ are obtained at the same heliocentric distance but in inbound and outbound legs of the orbit respectively. If, however, a different dependence of $R ( v_{impact} )$ is assumed, eq. \eqref{eq:velocity_first_estimate} changes. Assuming $R \propto v_{impact}^{q}$, a second estimate of $v_{dust;rad}$ is obtained by

\vspace{-20pt}

\begin{subequations}
\begin{align}
\begin{split}
    v_{dust;rad} &= |v_{sc;rad}| \frac{ \left( R_{in}^{2/q} + R_{out}^{2/q} \right)}{\left( R_{in}^{2/q} - R_{out}^{2/q} \right)}  +  \frac{\sqrt{\tilde{D}}}{\left( R_{in}^{2/q} - R_{out}^{2/q} \right)}, 
    \label{eq:velocity_second_estimate}
\end{split}\\
\begin{split}
    \tilde{D} &= v_{sc;rad}^{2} \left(R_{in}^{2/q} + R_{out}^{2/q} \right)^2 - v_{sc}^{2} \left(R_{in}^{2/q} - R_{out}^{2/q}\right)^2,
    \label{eq:velocity_second_estimate_discriminant}
\end{split}
\end{align}
\end{subequations}
where $q$ is equivalent to $1+\alpha \delta$ in eq. \eqref{eq:z2_model} and $\tilde{D}$ has no direct physical interpretation. We took the spacecraft's azimuthal velocity into account, but not the dust's azimuthal velocity, as that would be a second unknown component, for which we do not have enough information. It is nonetheless possible to correct for assumed dust azimuthal velocity by subtracting it from $v_{sc}$, see appendix \ref{app:velocity} for derivation of eqs. \eqref{eq:velocity_first_estimate} -- \eqref{eq:velocity_second_estimate_discriminant}.

It follows from eq. \eqref{eq:velocity_second_estimate} that having $R_{in}$, $R_{out}$ observed, the value of $q>1$ will lead to higher velocity estimate than in the case of $q=1$, an estimate that is higher by a factor of $q$ in first order approximation. \cite{zaslavsky2021first} reported inferred velocities $v_{dust;rad} \approx \SI{50}{km/s}$ assuming $q=1$ and shown compatibility of detection rates with the model assuming $q=1+\alpha \delta \approx 2.3$ according to eq. \eqref{eq:z2_model}. Assumptions met, $v_{dust;rad} \approx \SI{50}{km/s}$ is likely an underestimate. The most important assumption here is that the dust indeed comes from a hyperbolic population.

Note that the assumption that all detected dust grains are hyperbolic is difficult to verify or falsify. The most prominent trend in detections is that the counts diminish with increasing heliocentric distance, which could easily hide a plethora of other components, like bound dust or interstellar dust. First correction to the assumption that all detections come from hyperbolic dust stream is the assumption of having a two component field: hyperbolic dust and sporadic (background) detections, the latter having no dependence on spacecraft location nor velocity. 

\subsection{Velocity inference}

\begin{figure}[h]
\centering
\includegraphics{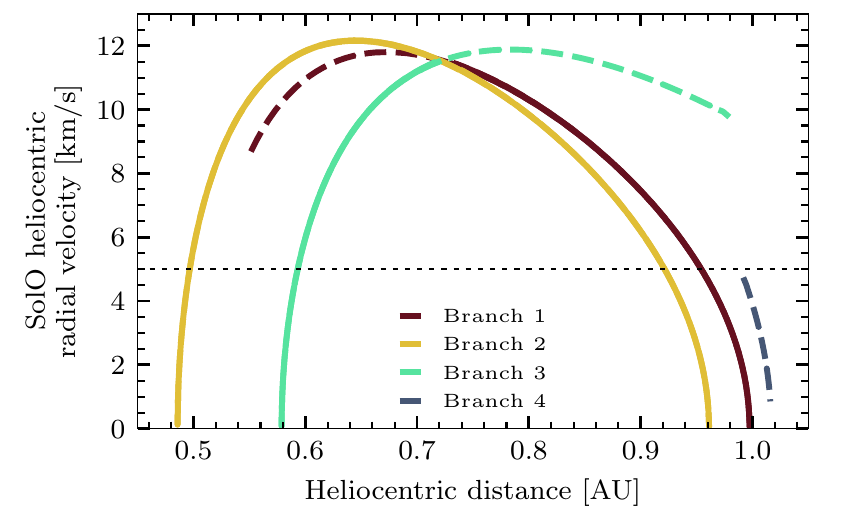}
\caption{SolO heliocentric distance and absolute value of its radial velocity. Colors separate individual branches of the orbit that come with changes in orbital parameters at gravity assists. Dashed lines correspond to all the combinations of radial velocity and location, while solid lines denote that SolO passed through both inbound and outbound arms for the combination. The horizontal dashed line denotes SolO radial velocity of $\SI{5}{km/s}$.}
\label{fig:plt:state_space}
\end{figure}

Assuming the dust flux is not explicitly dependent on time, the dust detection rate is a function of orbital phase as long as the orbital parameters don't change. Conversely, gravity assists change orbital parameters, such as perihelion, aphelion, and eccentricity. In the present analysis, we therefore treat sets of orbits, delimited by gravity assists, as separate data sets. In this way, data is aggregated for several orbits with the same orbital parameters, but we do not aggregate incompatible measurements. For instance: dust detection counts recorded near $\SI{0.6}{AU}$ on branches 2 and 3 are expected to be very different due to vastly different SolO radial velocity, see SolO radial velocity and its heliocentric location throughout its trajectory in Fig. \ref{fig:plt:state_space}. Minor orbital alterations between gravity assists are neglected. 

Since 29th June 2020, SolO has undergone 3 gravity assists, producing 4 distinct sets of data. The chronologically last of these so far did not accumulate data sufficient for analysis, and crucially did not produce any detections in the inbound part of the orbit at the time of analysis, hence the first 3 branches were used. The difference between detection rate in inbound and outbound leg of an orbit could be used for dust radial velocity inference, provided that radial spacecraft velocity is not negligible, compared to dust radial speed. Hence data represented by dashed lines in Fig. \ref{fig:plt:state_space} are not used for this analysis. Data with radial spacecraft velocity $>\SI{5}{km/s}$ are not used as they carry little information (see horizontal dashed line in Fig. \ref{fig:plt:state_space}). 

\begin{figure}[h]
\centering
\includegraphics{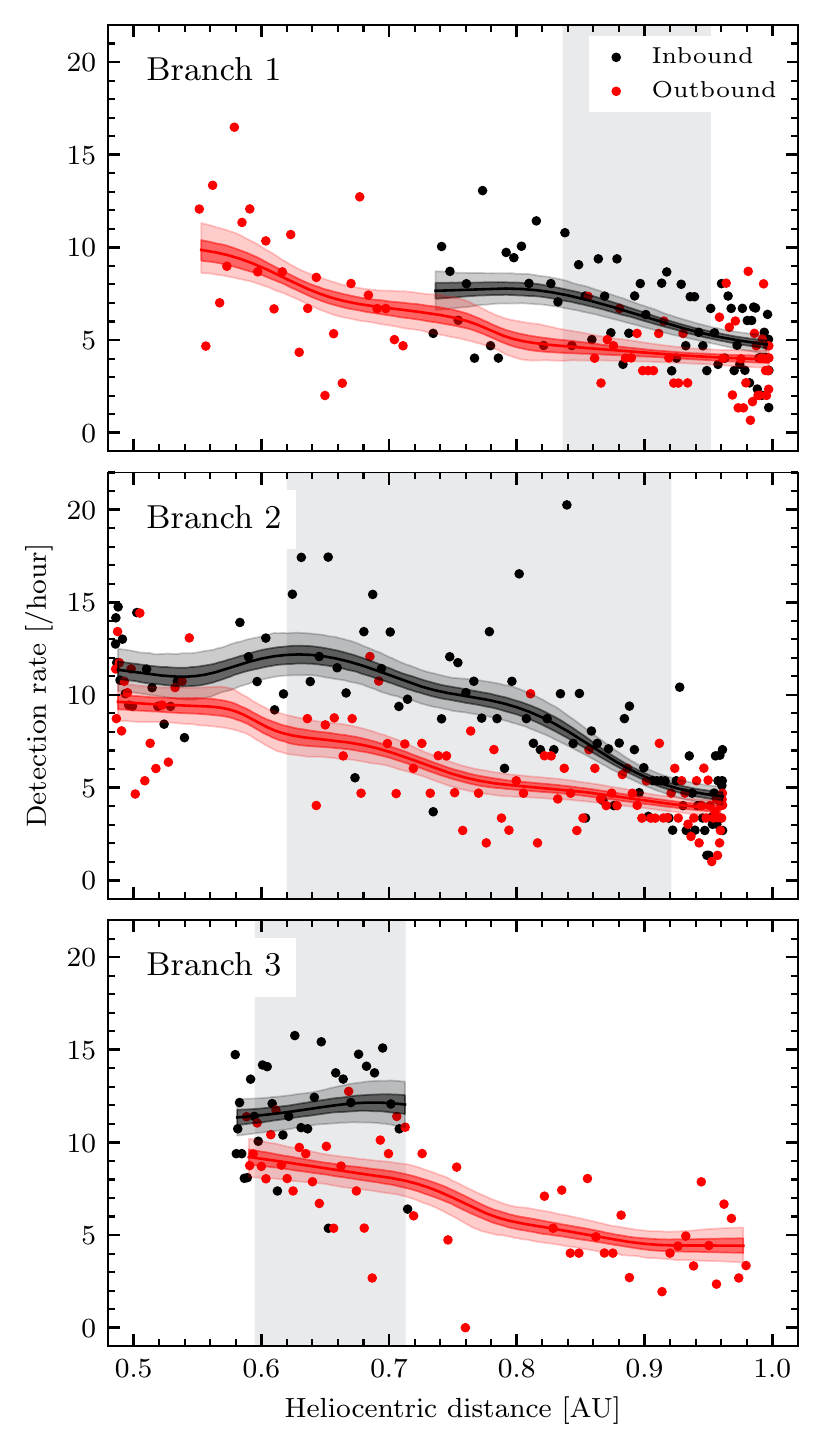}
\caption{Non-parametric fitting of detection rate observed between 29th JUN 2020 and 27th NOV 2021 in inbound and outbound part of the trajectory, branches separated by gravity assists on 26th DEC 2020 and 8th AUG 2021. The lines are the results of non-parametric fitting and only grayed intervals are used for further analysis, compare with Fig. \ref{fig:plt:state_space}.}
\label{fig:plt:nonparam_fit_all}
\end{figure}

In order to estimate radially dependent velocity, we produced smooth estimates of radially dependent detection rates, as defined in TDS/TSWF-E/CNN data set. The fitting was done separately for inbound and outbound legs, separately for each gravity assist delimited data set. In order to not rely on assumptions, we decided to use non-parametric fitting, specifically Nadaraya–Watson kernel regression \citep{watson1964smooth,nadaraya1964estimating} with a Gaussian kernel ($FWHM \approx 2.355\sigma = \SI{0.15}{AU}$). This is a simple and robust local-averaging fitting procedure, producing $C^\infty$ estimates. Dust detection counts are Poisson random variables, therefore they have variance equal to their mean value. To evaluate uncertainty, we constructed confidence intervals for the non-parametric fit by bootstrapping on daily dust counts: new samples were generated with original counts as rates for new Poisson-distributed random variables. For illustration of all three data sets and fitted rates, see Fig. \ref{fig:plt:nonparam_fit_all}. For every branch, we only used the heliocentric distance interval where both inbound and outbound legs are available, bounded by the innermost and the outermost detection on the leg, see the grayed areas in Fig. \ref{fig:plt:nonparam_fit_all}. We did not use the $r<\SI{0.62}{AU}$ of branch 2, as there are no outbound detections near $\SI{0.6}{AU}$ and the detections near $\SI{0.5}{AU}$ show little difference between the inbound and the outbound leg. This may be due to spatial limitation of the given model, an unlikely combination due to scarce data, a truly higher radial velocity, or a combination of more effects. 

\begin{figure}[h]
\centering
\includegraphics{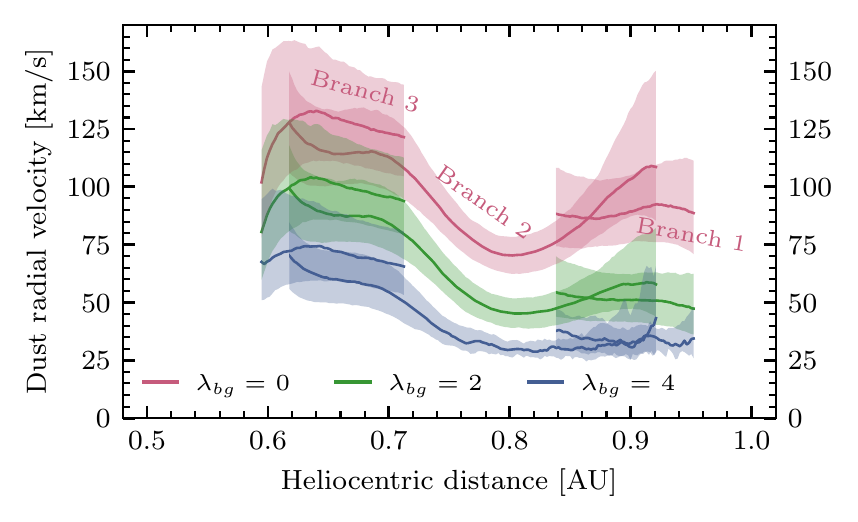}
\caption{Velocity estimated from TDS/TSWF-E/CNN. Different colors correspond to different assumed background rates. Shades correspond to $\SI{50}{\%}$ confidence intervals and the solid lines correspond to median values for a given heliocentric distance. Different branches of $\lambda_{bg} = \SI{0}{h^{-1}}$ are labeled. Only parts of the branches where $v_{SolO}>\SI{5}{km/s}$ are shown.}
\label{fig:plt:velocity_profile_only_background}
\end{figure}

Having smooth detection rate estimates, we produced velocity estimates using eq. \eqref{eq:velocity_second_estimate}, see Fig. \ref{fig:plt:velocity_profile_only_background}. Bootstrap samples of detection rates are used to calculate the shown percentile confidence intervals. Notably, not all the bootstrap samples for $\lambda_{bg}=4$ allowed for a solution, which is apparent from the jitters of the blue curve at $r > \SI{0.75}{AU}$. Confidence intervals are constructed from the solutions that were obtained. This issue is to be expected, as $\lambda_{bg}=4$ implies very little hyperbolic dust at $r > \SI{0.8}{AU}$ (see Fig. \ref{fig:plt:nonparam_fit_all}) and therefore uncertainty in the inferred velocity. The estimate shown in Fig. \ref{fig:plt:velocity_profile_only_background} assumes $\alpha \delta = 1.3$ and an initial heliocentric distance of $\SI{0.1}{AU}$, the latter in the form of correction for dust azimuthal velocity. 

To further estimate uncertainty, we included three relevant parameters (in total): 

\begin{itemize}
    \item a background (non-hyperbolic) rate $\lambda_{bg}$, corrected for by subtraction from the estimated detection rate,
    \item the product $\alpha \delta$ included in eq. \eqref{eq:velocity_second_estimate},
    \item azimuthal dust velocity corresponding to different initial circular orbits, as shown in Fig. \ref{fig:plt:beta_azimuthal_velocity}, giving a straight-forward generalization of eq. \eqref{eq:velocity_second_estimate}.
\end{itemize}

\begin{figure}[h]
\centering
\includegraphics{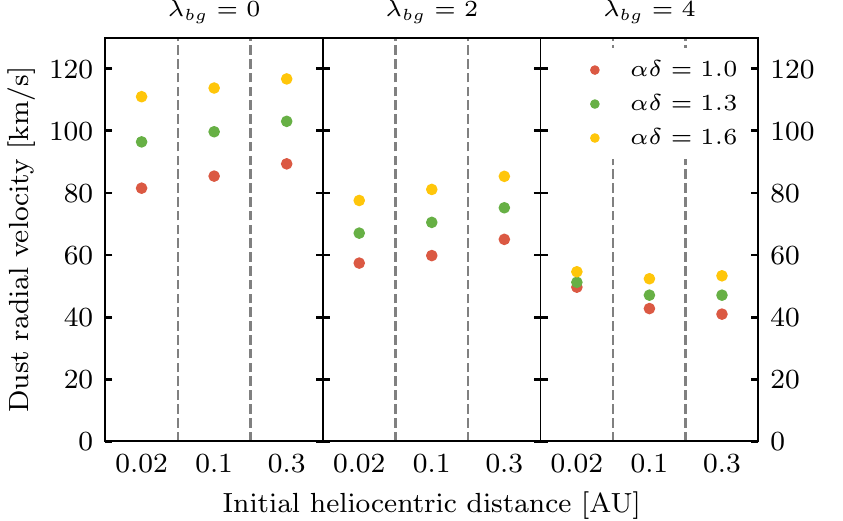}
\caption{Velocity estimated from TDS/TSWF-E/CNN. Single velocities are obtained from profiles for better readability. The points are constructed as averages of velocities at $0.65, 0.75, 0.85 \si{AU}$ for all the branches, where a velocity is the median solution for all the bootstrap replications. For full profiles see appendix \ref{app:full_velocity_profiles}.}
\label{fig:plt:velocity_other_parameters}
\end{figure}

We have an estimate of the region of likely velocities (see Fig. \ref{fig:plt:velocity_other_parameters}), given reasonable variation of free parameters. The analysis shows the velocity to be mostly between ${40}{km/s}$ and ${100}{km/s}$ and the rate of constant, non-hyperbolic dust between $0 - 4 \si{\per\hour}$ (according to Fig. \ref{fig:plt:velocity_other_parameters}). The background detection rate $\lambda_{bg}$ could clearly not be lower than $0$ and the result suggests it could not be much higher than $\approx \SI{4}{h^{-1}}$, because no solutions of \eqref{eq:velocity_second_estimate} in that case, as missing solutions in Fig. \ref{fig:plt:velocity_profile_only_background} show. The reason is, that the difference between inbound and outbound rates are observed to be too high to be explained in the case of $\lambda_{bg} = \SI{4}{h^{-1}} $. A rather low amount ($\lesssim \SI{1}{h^{-1}}$) of non-hyperbolic dust would imply higher velocity, in the range of $\approx \SI{100}{km/s}$. The conclusion is that the higher the background detection rate $\lambda_{bg}$ is, the lower is the underlying dust velocity. Similarly, a higher $\alpha \delta$ product implies higher velocity and larger initial radius (in case of $\beta$-meteoroids) implies higher underlying radial velocity. Furthermore, assuming $\beta$-meteoroids, low velocities $\gtrsim \SI{50}{km/s}$ imply a low $\beta$ factor (see Fig. \ref{fig:plt:beta_rad_velocity}). While bearing many uncertainties, this inference is very robust as it does not depend on a specific model for dust, in particular it is independent on dust spatial density as a function of heliocentric distance, because we only compare observations on the same heliocentric distance. The background component is among the biggest unknowns, see appendix \ref{app:full_velocity_profiles} for full velocity profiles that produce the data in Fig. \ref{fig:plt:velocity_other_parameters}.

\subsection{Spatial density}  \label{ch:spatial_density}

If the hyperbolic grains do not accelerate (for example $\beta$-meteoroids with $\beta \approx 1$ are assumed), a radial dependence of spatial density of $\sim r^{-2}$ is the result. This is not the case if acceleration or deceleration is present. Particularly, it makes sense to assume slowing dust ($\beta < 1$), as figure \ref{fig:plt:velocity_profile_only_background} suggests slowing rather than accelerating dust. Also, $\beta \approx 1$ or even $\beta > 1$ needs a rather specific set of conditions (combination of material and specific size, see \cite{mann2010interstellar}), while $0.5 < \beta < 1$ is possible for a broad range of dust parameters. The observed effective $\beta$ is then determined by aggregation of all components. The equation for detection rate \eqref{eq:z2_model} contains $r^{-2}$, but it remains the correct expression for dust flux even if $v_{dust}$ is not constant. In that case, $r^{-2}$ should not be interpreted as a spatial density of dust, but as a geometric factor. The spatial density is then expressed through the non-constant $v_{dust}$. However, this makes it very difficult to fit model (eq. \eqref{eq:z2_model}) to the data, as $v_{dust}$ is no longer a numeric parameter, but a function of $r$.

\begin{figure}[h]
\centering
\includegraphics{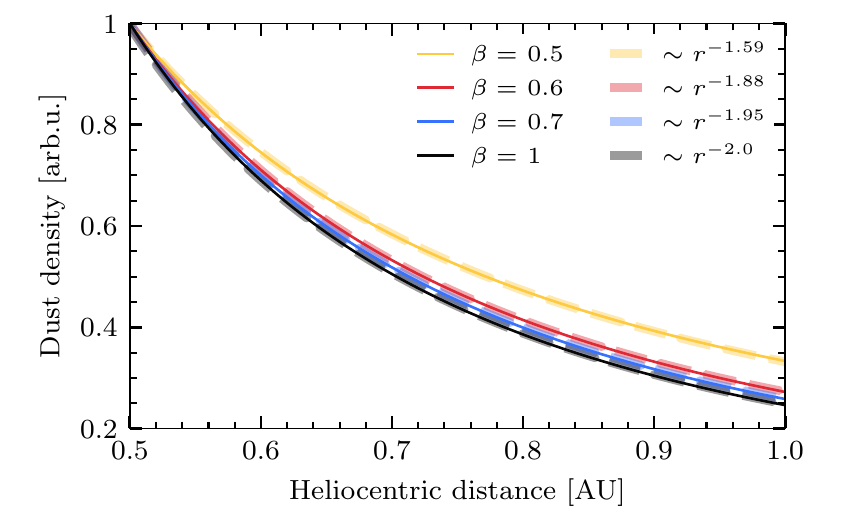}
\caption{Modelled dust spatial densities for different $\beta$ values assuming a circular initial orbit of $\SI{0.1}{AU}$. Solid lines show the spatial density and are normalized to the density at $\SI{0.5}{AU}$. Dashed lines are approximations to the solid lines, assuming power dependence on $r$.}
\label{fig:plt:distance_exponents}
\end{figure}

Let us stick to spatial density view and examine the effective exponent of $r$ given $\beta$-meteoroids with some $0.5 \leq \beta < 1$. Fig. \ref{fig:plt:distance_exponents} shows an example of how the $\beta$ value influences the spatial dust density (for spatial dust density calculation, see appendix \ref{app:single_particle_valocities}). The analysis of spatial density as a result of deceleration does not require the dust to be $\beta$-meteoroids, but the relation to $\beta$ value clearly does. In Fig. \ref{fig:plt:distance_exponents}, an initial orbit of $\SI{0.1}{AU}$ is assumed, see appendix \ref{app:distance_exponents_different_initial} for plots of dust spatial density variation similar to Fig. \ref{fig:plt:distance_exponents} for different initial orbits. The particular exponents depend on the initial orbit, but the general trend of lower $\beta$ value implying deceleration remains. 

\section{Daily count inference} \label{ch:inla}

\subsection{Model formulation} \label{ch:model_formulation}

We decided to model the number of dust detections within a day as a Poisson distributed random variable, as dust detection itself is a prime example of a Poisson point process in time, as discussed in section \ref{ch:poisson}. The rate $\lambda$ of the process is considered dependent on multiple parameters $\vec{\theta}$. Notably, we consider $\lambda$ to not be explicitly dependent on time, but to be temporarily dependent indirectly, through the orbital parameters and the orbital phase of the spacecraft. Importantly, the rate is also considered dependent on the parameters of the dust cloud. Therefore, we formulate a hierarchical Bayesian model with $5$ parameters: $\epsilon_v, \epsilon_r, \lambda_{\beta}, \lambda_{bg}, \nu_r$, that we for simplicity denote $\vec{\theta} = (\epsilon_v, \epsilon_r, \lambda_{\beta}, \lambda_{bg}, \nu_r)$. These parameters are used to model the rate $\lambda$ and by extension detected counts, see equations \eqref{eq:model_likelihood} to \eqref{eq:model_dust_velocity}. Note that the rate $\lambda$ (see eq. \eqref{eq:model_rate}) is a generalisation of eq. \eqref{eq:z2_model} with an additional constant (background) term and a variable exponent of heliocentric distance. In order to keep the present model simple and yet allow for non-constant velocity, we use the parameters $\nu_r$ and $\epsilon_r$ as mean radial dust velocity and density exponent respectively, see appendix \ref{app:epsilon_r} for further interpretation. This combination allows for fitting of a single constant mean velocity and effective acceleration (through the exponent $\epsilon_r$) at the same time with just two constant scalar parameters. Mind though, that the parameter $\nu_r$ has the exact meaning of mean velocity only in the case of constant velocity. Given our generalization, it is not necessarily the mean velocity over the orbit, but an effective velocity parameter. This is acceptable, as the velocity is reasonably approximated by a constant between $\SI{0.5}{AU}$ and $\SI{1}{AU}$, even in the case of $\beta = 0.5$ (see Fig. \ref{fig:plt:beta_rad_velocity}).  

The Poisson likelihood (eq. \eqref{eq:model_likelihood}) includes exposure time $E$ (in hours) and the rate $\lambda$ (in detections per hour). Then eq. \eqref{eq:model_relative_velocity} is a straight-forward definition of relative velocity between the spacecraft and the dust particle, while eq. \eqref{eq:model_dust_velocity} describes the decomposition of dust velocity into radial and azimuthal components. In practice, the model defines the parameter $\nu_r$ as the radial velocity of a dust particle and the variable $\nu_{azimuthal}$ as the azimuthal velocity of the same particle, but only $\nu_r$ is regarded as a random variable. The variable $v_{a}$ is directly related to heliocentric distance $| \vec{r} |$ according to eq. \eqref{eq:model_azimuthal_velocity}, which is approximately equivalent to the $r_0 = \SI{0.1}{AU}$ line in Fig. \ref{fig:plt:beta_azimuthal_velocity}. This is due to the simpler and less important dependence of $v_{impact}$ on $v_{a}$ and as a compromise in order to keep the number of free parameters reasonable with respect to available data (the attempts to fit $6$ parameters were not fruitful). The main goal of the fitting procedure is to get the marginal posterior distributions of each of the parameters $\vec{\theta}$ of the model

\vspace{-20pt}

\begin{subequations}

\begin{align}
\begin{split} 
    \qquad \qquad N | \lambda, \vec{\theta}  &\sim Poiss(E \cdot \lambda(\theta) ),
    \label{eq:model_likelihood}
\end{split}\\
\begin{split}
    \lambda (\vec{\theta}) &= \lambda_{\beta} \cdot v_{impact}^{\epsilon_v} \cdot r^{\epsilon_r} + \lambda_{bg},
    \label{eq:model_rate}
\end{split}\\
\begin{split}
    v_{impact} &= \frac{| \vec{v_{SolO}} - \vec{v_{dust}} |}{\SI{50}{km/s}},
    \label{eq:model_relative_velocity}
\end{split}\\
\begin{split}
    \vec{v_{dust}} &= \nu_r \cdot \vec{e_r} + v_{a} \cdot \vec{e_\phi},  
    \label{eq:model_dust_velocity}
\end{split}\\
\begin{split}
    v_{a} &= \SI{12}{km/s} \, \frac{\SI{0.75}{AU}}{| \vec{r} |}.
    \label{eq:model_azimuthal_velocity}
\end{split}
\end{align}
\end{subequations}

\vspace{10pt}

There are $N$ detections observed in a given day, the exposure $E$ is known, and so are the location and velocity of SolO. In equation \eqref{eq:model_relative_velocity}, a dimensionless parameter is constructed --- it has computational advantages if $v_{impact} \approx 1$, as rather high power of the variable is computed in the process. Eqs. \eqref{eq:model_relative_velocity} to \eqref{eq:model_azimuthal_velocity} explain the role of the parameter $v_r$ in eq. \eqref{eq:model_rate} and are only separated from \eqref{eq:model_rate} for better readability. Note that purely 2D motion of dust particles, within the ecliptic plane, is assumed in eq. \eqref{eq:model_dust_velocity}.

The parameter $\epsilon_v$ is the exponent of $v_{impact}$ in the mean rate formula and incorporates the dependence on rate of volume scanning ($V/t \propto S \cdot v_{impact}$), hence $v_{impact}^1$, and the dependence on charge yield $\alpha$ and dust mass power-law exponent $\delta$ in the form of $v^{\alpha \delta}$. The dependence is then $v_{impact}^{1 + \alpha \delta} = v_{impact}^{\epsilon_v}$.

The parameter $\epsilon_r$ is the exponent of heliocentric distance $r$, notably influenced by acceleration / deceleration of dust, as discussed in sec. \ref{ch:spatial_density}. See appendix \ref{app:epsilon_r} for further interpretation.

The parameter $\lambda_\beta$ plays the role of a normalization constant, accounting for an absolute dust spatial density and spacecraft detection area and holds the physical unit of $hour^{-1}$. It is uninteresting to study this parameter in itself, in the sense that it merely normalizes the model so that the detection rate correspond to the observed mean rate and has no consequence on the physical characteristics of any given particle.

The parameter $\lambda_{bg}$ has the meaning of detections per hour as well, but has a clearer interpretation as the background detection rate, that is the rate of detections that are not attributable to hyperbolic dust.

The parameter $\nu_r$ also has a very direct meaning, which is the mean outward radial velocity of the hyperbolic dust in our experimental range. Note that variation of impact velocity is still allowed by variation in spacecraft velocity $\vec{v_{SolO}}$. Acceleration is accounted for in $\epsilon_r$.

\subsection{Prior distributions of parameters}

For Bayesian inference, choosing reasonable priors is important. Ideally, priors should be informative (narrow) enough to capture the prior knowledge about parameters, but vague (wide) enough so that they still allow for additional information to play a role. It is physically infeasible for the parameters $\lambda_\beta$ and $\lambda_{bg}$ to be negative, as they have a meaning of detection rate. Furthermore, positive radial velocity is also required by the model to work. Therefore, we opted for gamma priors for these three parameters. Although we are quite sure about the sign of the parameters $\epsilon_v$, $\epsilon_r$, neither the model nor the physical unit actually rules out the possibility of $\epsilon_v$, $\epsilon_r$ having any sign. We therefore opted for normal priors for $\epsilon_v$, $\epsilon_r$. The choice of prior family for $\epsilon_v$, $\epsilon_r$, and $\nu_r$ is of little importance. Generally speaking, prior choice makes less of a difference the more data are analyzed. 

\begin{figure}[h]
\centering
\includegraphics{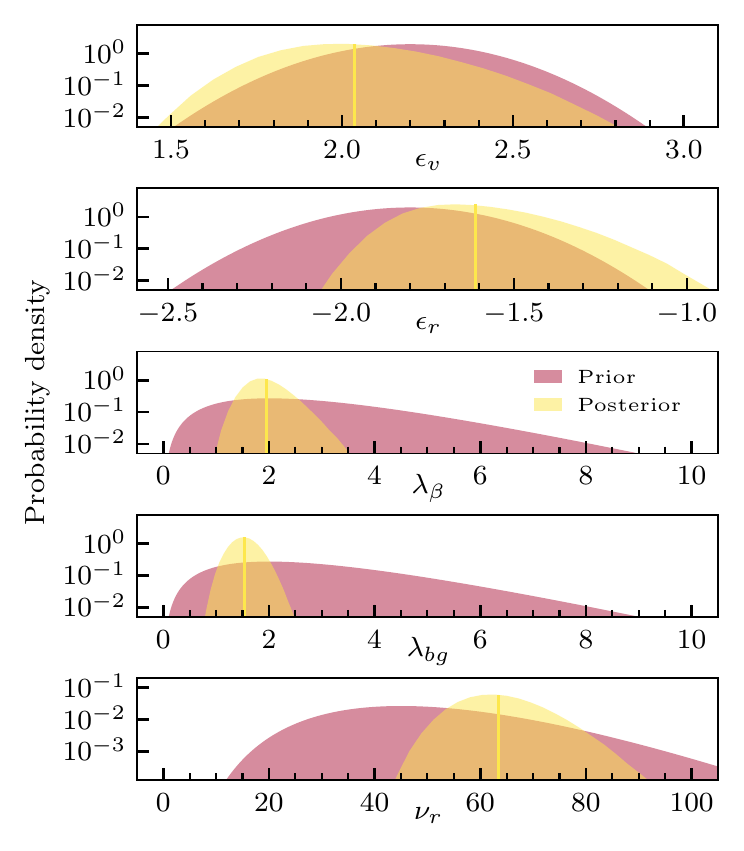}
\caption{Prior and posterior distributions for the parameters $\theta$. The prior distributions are described in text. Summary statistics for posterior distributions are described in Tab. \ref{tab:posteriors}.}
\label{fig:plt:priors_posteriors}
\end{figure}

In order to incorporate our actual prior belief about the model, we chose what we believe are moderately informative priors for the parameters. The following paragraphs discuss our choice. For a graphical representation of the prior distributions of the parameters, see Fig. \ref{fig:plt:priors_posteriors}. 

The parameter $\epsilon_v$ stands for $1 + \alpha \delta$. Since we have indications from \cite{zaslavsky2021first} that $\delta \approx 0.3$ and most laboratory experiments show \citep{collette2014micrometeoroid} that $3 \lesssim \alpha \lesssim 5$, we expect $1.9 \lesssim \epsilon_v = 1 + \alpha \delta \lesssim 2.5$. We therefore choose the prior $\epsilon_v \sim Norm(mean = 2.2, stdev = 0.2)$, which places emphasis on the range $2.0 < \epsilon_v < 2.4$ and yet doesn't prohibit any real $\epsilon_v$. 

Provided that there are no major sources of dust between $\SI{0.5}{AU}$ and $\SI{1}{AU}$ and provided that dust does not either accelerate nor decelerate, $\epsilon_r = -2$, which follows easily from mass conservation. If we relax the latter assumption, then $\epsilon_r \neq -2$. In fact, the dependence will no longer follow $r^{\epsilon_r}$ exactly, but as is shown in Fig. \ref{fig:plt:distance_exponents}, for $\beta$-meteoroids of $0.5 \lesssim \beta \lesssim 1$ the dependence is very similar to $r^{\epsilon_r}$ with $-2 \lesssim \epsilon_r \lesssim -1.59$. We therefore chose a prior $\epsilon_r \sim Norm(mean = -1.8, stdev = 0.2)$ which emphasizes the range $-2.0 < \epsilon_r < -1.6$ but in principle allows for any real $\epsilon_r$. 
As for the parameter $\lambda_\beta$, we know it is on the order of the total rate, which is on average $\SI{6.9}{h^{-1}}$. The interpretation of the parameter is made more opaque by the normalization in eq. \eqref{eq:model_relative_velocity}. However, the factor of $v_{impact}^{\epsilon_v}$ is on the order of $1$ and the factor of $r^{\epsilon_r}$ is $>1$, hence we expect $1 \lesssim \lambda_\beta \lesssim 10$. We chose a less informative prior of $\lambda_\beta \sim Gamma(shape = 3, scale = 1)$. 

Fig. \ref{fig:plt:velocity_profile_only_background} showed that for background detections, $\lambda_{bg} < \SI{4}{hour^{-1}}$ is feasible. We chose a less informative prior $\lambda_{bg} \sim Gamma(shape = 3, scale = 1)$, which is wide and allows for any positive $\lambda_{bg}$. 

Based on Fig. \ref{fig:plt:velocity_other_parameters}, we believe that values $\SI{40}{km/s} \lesssim \nu_r \lesssim \SI{80}{km/s}$ are mostly expected. We chose the prior $\nu_r \sim Gamma(shape = 10, scale = 5)$ that emphasizes that range, with the mean of $\SI{50}{km/s}$, which is the value that \cite{zaslavsky2021first} reported. This prior still allows for any positive value of $\nu_r$. 

\subsection{Posterior distributions} \label{ch:posterior_distributions}

The analysis is performed using TDS/TSWF-E/CNN data. For an analogous analysis performed on SolO on-board identified dust impacts, see appendix \ref{app:tds_inla}. Posteriors are inferred using R-INLA. The model \eqref{eq:model_rate} is complicated by the steep dependence of the rate $\lambda$, especially the dependence on exponential parameters $\epsilon_v, \epsilon_r$. Note that a radial velocity $\gtrsim \SI{60}{km/s}$ is consistent with detection rate $\lambda_{bg} \approx 1.5$ and $\alpha \delta \approx 1.0$, according to Fig. \ref{fig:plt:velocity_other_parameters}. It is important to note that the exact choice of priors and other parameters, such as the reference azimuthal velocity in eq. \eqref{eq:model_relative_velocity} and procedure starting point (as INLA works on a grid largely defined by the initial point) influences the exact result, although no major difference is encountered when parameters or priors are reasonably varied (see appendix \ref{app:priors_variation}). As for the initial point, the mode of the joint prior was used: $\vec{\theta} = (2.2,-1.8,2,2,45)$. 

Several measures can be used to evaluate the appropriateness of a model to a data set. We inspected the conditional predictive ordinates (CPO) and the predictive integral transform (PIT), that indicated no issues, see appendix \ref{app:model_evaluation} for details. 

\begin{table}[]
\centering
\begin{tabular}{r|c|c}
   & $Mean$  & St. dev. \\ \hline
$\epsilon_v$ & 2.04  & 0.20 \\ \hline
$\epsilon_r$ & -1.61 & 0.16 \\ \hline
$\lambda_\beta$ & 1.96  & 0.38 \\ \hline
$\lambda_{bg}$ & 1.54  & 0.25 \\ \hline
$\nu_r$ & 63.4  & 6.7 
\end{tabular}
\caption{The marginal posterior mean and the standard deviation for all the parameters, see Fig. \ref{fig:plt:priors_posteriors} for visual representation of the posterior distributions.}
\label{tab:posteriors}
\end{table}

\subsection{Discussion of the posterior distribution} \label{ch:posterior_distribution}

The inferred posterior distribution of velocities shown in Fig. \ref{fig:plt:priors_posteriors} is not to be interpreted as a distribution of velocities within the dust cloud directly, but rather as a distribution of the effective mean velocities encountered on each day, or even better --- the uncertainty in effective velocity. There could indeed be dust grains with velocity well off the effective support of the posterior distribution, as long as the mean of all velocities does not exceed the region indicated by the posterior distribution. 

\begin{figure}[h]
\centering
\includegraphics{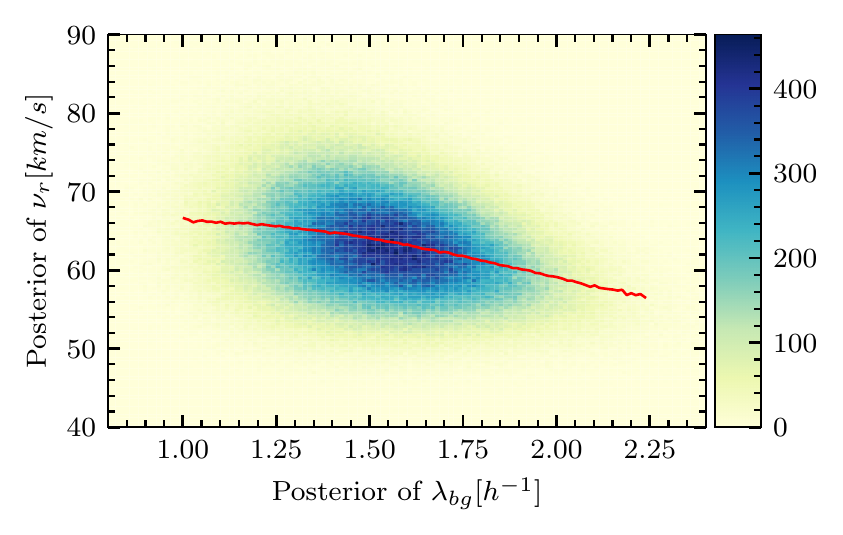}
\caption{The covariance between $\nu_r$ (radial dust velocity) and $\lambda_{bg}$ (background detection rate), correlation is $-0.3$. The red line is the mean $\nu_r$ conditioned on $\lambda_{bg}$. Produced by sampling from the joint posterior distribution of $\vec{\theta}$.}
\label{fig:plt:covariance_c2_v1}
\end{figure}

The parameters $\vec{\theta}$ are not independent. Fig. \ref{fig:plt:covariance_c2_v1} shows the covariance between the parameters $\nu_r$ (radial velocity) and $\lambda_{bg}$ (background detection rate). Negative correlation suggests that higher $\nu_r$ is likely in case of lower $\lambda_{bg}$. This offers a sanity check: higher velocity would mean lower difference between inbound and outbound flux, which has a similar effect to the higher background component scenario --- negative correlation between $\nu_r$ and $\lambda_{bg}$ is thus expected. For covariances between all parameters, see appendix \ref{app:covariance_posteriors}.

The TDS/TSWF-E/CNN data set contains $\SI{6.9}{h^{-1}}$ detections on average. The inferred value of $\lambda_{bg} = \SI[separate-uncertainty]{1.54 \pm 0.25}{h^{-1}}$ implies that in total $\SI[separate-uncertainty]{78 \pm 4}{\%}$ of dust is attributed to hyperbolic dust within the model. The constant background $\lambda_{bg}$ is the simplest available generalization and is therefore likely an oversimplification. Hyperbolic dust detection rate shows strong negative correlation with heliocentric distance. If, for instance, non-hyperbolic dust shows similar anti-correlation, the actual non-hyperbolic component is higher than inferred. Conversely, if non-hyperbolic prevalence shows correlation with heliocentric distance, the actual non-hyperbolic component is lower than inferred. Both cases would also imply changes to the inferred parameters of hyperbolic dust, see appendix \ref{app:background} for a visual explanation. If non-hyperbolic is mostly non-dust (for example misattributed electrical phenomena), independence on heliocentric distance is reasonable. However, if most non-hyperbolic are bound (Keplerian) dust particles, then anti-correlation is expected. If interstellar dust (ISD) streaming predominantly from one direction approximately within the ecliptic plane is present, positive correlation is also feasible due to velocity vector orientation. 

Indeed, ISD was observed \citep{baguhl1996situ,zaslavsky2012interplanetary,malaspina2014interplanetary} to arrive mainly from $\SI{258}{^\circ{}}$ ecliptic longitude. The highest flux is observed when a spacecraft has anti-parallel velocity, which vaguely coincides with higher heliocentric distance phase of SolO's orbit so far. If ISD is an important contribution to $\lambda_{bg}$, the actual background flux may be lower than the suggested $\lambda_{bg} \approx \SI{1.5}{h^{-1}}$. For now, ISD is not apparent in SolO data and the fact that models fit well without ISD suggests it is not an important component of SolO detections. Its identification with SolO is however beyond the scope of the present work, but remains worthy of future investigation, especially since ISD may become more important during the current solar cycle \citep{mann2010interstellar}. For now, no bound dust particles are apparent either, nor are the retrograde dust particles. If the constant background is a crude oversimplification and the non-hyperbolic component has a prominent dependence on heliocentric distance, the present interpretation of the parameters $\vec{\theta}$ is not correct, as the model is not on point. Inclusion of more parameters in the model (for example a more sophisticated non-hyperbolic term) may be feasible with more data in coming months. 

\begin{figure}[h]
\centering
\includegraphics{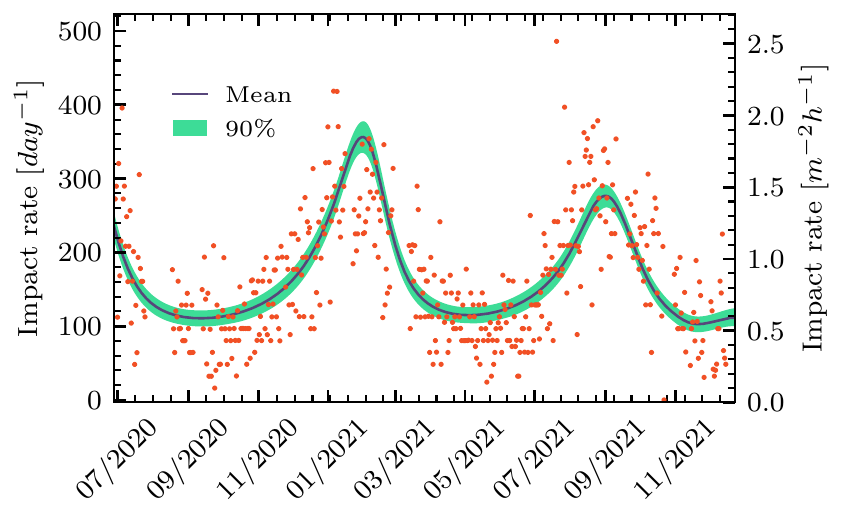}
\caption{The estimated posterior mean of the dust impact with $\SI{90}{\%}$ HPD credible intervals. The credible intervals are not supposed to cover the data scatter (see text for interpretation of shown credible intervals).}
\label{fig:plt:modelled_flux_rate}
\end{figure}

\begin{figure}[h]
\centering
\includegraphics{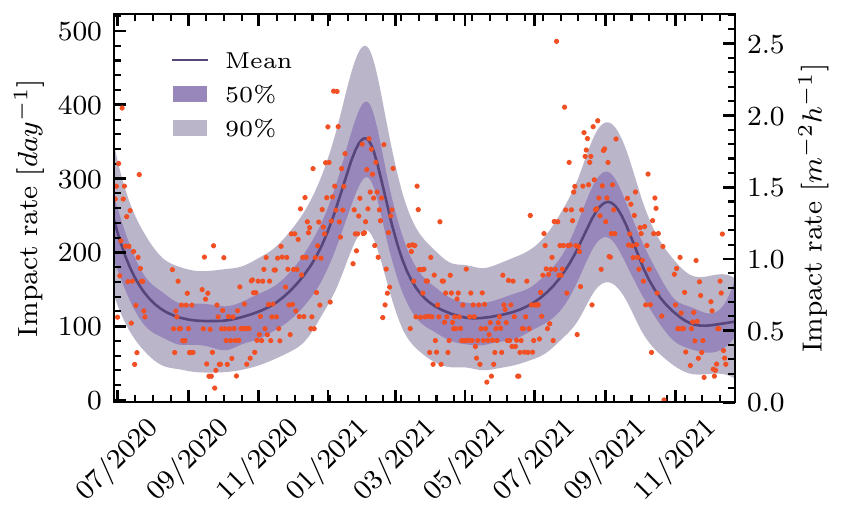}
\caption{The estimated posterior mean of the hyperbolic dust detections and HPD prediction intervals. The Prediction intervals are supposed to cover the data scatter (see text for interpretation of shown prediction intervals).}
\label{fig:plt:modelled_flux}
\end{figure}

The posterior mean of the detection rate is shown in Fig. \ref{fig:plt:modelled_flux_rate} in units of: $\si{m^{-2} h^{-1}}$, assuming a detection area of $\SI{8}{m^2}$ (SolO thermal shield approx. area); and $\si{day^{-1}}$, taking into account the detection time per day and extrapolating to $\SI{24}{h}$. Note that the credible intervals reflect the uncertainty of the inferred mean detection rate (the uncertainty of our knowledge, given the data), which is the same uncertainty as visualised in Fig. \ref{fig:plt:priors_posteriors}. The spread of data points in Fig. \ref{fig:plt:modelled_flux_rate} is much wider and mostly defined by the variance of Poisson random variable, given the mean rate, rather than the uncertainty in the mean rate. The prediction intervals of the Poisson random variable are shown in Fig. \ref{fig:plt:modelled_flux} and there, data points seem to be appropriately covered by the credible intervals. 

\begin{figure}[h]
\centering
\includegraphics{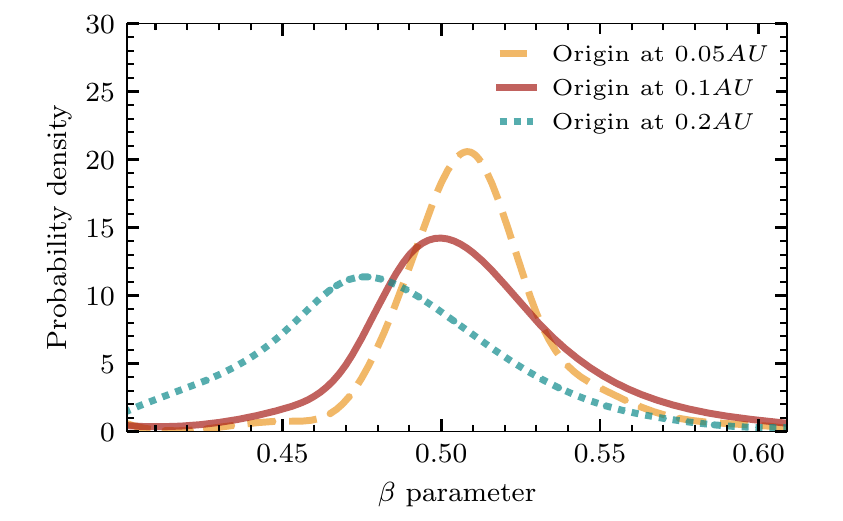}
\caption{$\beta$ parameter resulting from $\epsilon_v$ posterior distribution under the assumption of circular parent body orbit orbit (see text for discussion).}
\label{fig:plt:beta_hist}
\end{figure}

The inferred value of the parameter $\epsilon_r \approx -1.6$ suggests that dust grains are slowing substantially on their way out of the inner heliosphere between $\SI{0.5}{AU}$ and $\SI{1}{AU}$, resulting in a spatial distribution different from trivial $\lambda \propto r^{-2}$ case. See Fig. \ref{fig:plt:distance_exponents} for comparison. With inferred velocity of $\SI[separate-uncertainty]{63 \pm 7}{km/s}$ between $\SI{0.5}{AU}$ and $\SI{1}{AU}$, significant deceleration suggest much higher velocity closer to the Sun. Assuming $\beta$-meteoroids with a circular initial orbit, the $\epsilon_r$ value implies a specific $\beta$ value needed for just the right level of deceleration. Deceleration is a result of energy transfer from kinetic to potential, and therefore, given initial heliocentric distance, the deceleration rate depends on initial velocity. This makes the assumption of circular initial orbit crucial when we are to infer the $\beta$ parameter. For example, a $\beta$ value needed to explain an observed $\epsilon_r$ is different if the $\beta$-meteoroid parent object has eccentricity $0.3$, rather than $0$. For analysis of implied $\beta$ values in the case of circular parent orbit, see Fig. \ref{fig:plt:beta_hist}. Various initial parent body orbit radii are shown in Fig. \ref{fig:plt:beta_hist} to demonstrate that the model is not very sensitive to that parameter. For comparison, note that velocities $\gtrsim \SI{60}{km/s}$ are consistent with $\beta \approx 0.6$ and origin between $\SI{0.05}{AU}$ and $\SI{0.1}{AU}$, according to Fig. \ref{fig:plt:beta_rad_velocity}. Note that $\SI{0.05}{AU} \approx \SI{10}{R_\odot}$, where $R_\odot$ is the Solar radius. 

However, it is feasible to expect a parent body with eccentricity of $0.3$, as the mean eccentricity in the inner asteroid belt is $e \approx 0.15$ \citep{malhotra2009eccentricity}. If a dust grain is ejected from a given heliocentric distance $r$, the eccentricity $e=0.3$ implies $+14\%$ ejection speed if $r$ is the perihelion and $-16\%$ ejection speed if $r$ is the aphelion, compared to ejection from circular orbit of radius $r$. In the case of $e\neq0$, $\beta > 0.5$ is not the right condition for unbound $\beta$-meteoroid. In fact, for $e=0.3$  the condition is approximately $\beta > 0.35$ for perihelion, and $\beta > 0.65$ for aphelion ejection. Mind that the $+14\%$ could also be $\Delta v$ transferred at collision, as collisions between larger dust objects are likely a major source of $\beta$-meteoroids. Then the $14\%$ relative speed would for instance correspond to collision of two asteroids on circular orbits with relative inclination of $\SI{8}{^\circ}$, which is also a very feasible scenario. For instance, if the example of $+14\%$ of $\Delta v$ (or eccentricity of $0.3$) is a good representative of the process, the resulting implied $\beta$ would be not $\beta \gtrsim 0.5$, but rather $\beta \gtrsim 0.35$. Eccentricities and relative velocities in zodiacal cloud remain uncertain. Note that even in the described case, we are still considering a dust grain with purely azimuthal velocity at liberation, which is yet another simplification. 

\subsection{Comparison with previous results}

As Solar Orbiter has been operated only since 2020 and will be operated at least until 2027, the results presented in this paper are one of the earlier for the mission. Based on similar data, though collected over shorter time period, \cite{zaslavsky2021first} reported several physical parameters of the $\beta$-meteoroid population. Interestingly, they reported radial velocity to be about $\SI{50}{km/s}$, which is within 2 standard deviations from $\SI[separate-uncertainty]{63 \pm 7}{km/s}$ reported here, but bear in mind that the number is inferred under substantially different assumptions. The velocity is crucial for inference of $\beta$-meteoroid flux at $\SI{1}{AU}$ for example, which \cite{zaslavsky2021first} reported to be $\SI{8e-5}{m^{-2}s^{-1}}$. Under our model assumptions (constant radial dust velocity) and taking joint posterior distribution of the parameters, we report $\SI[separate-uncertainty]{1.6(1) e-4}{m^{-2}s^{-1}}$ for hyperbolic dust and the residual component ($\lambda_{bg}$) together (measured on a stationary spherical object, per $m^2$ of cross section), a value higher by a factor of $\approx 2$. For the component consistent with hyperbolic dust only, we report the flux of $\SI[separate-uncertainty]{1.1(2) e-4}{m^{-2}s^{-1}}$ and for the component attributed to the residual, background component $\SI[separate-uncertainty]{5.4(15) e-5}{m^{-2}s^{-1}}$. As for $\alpha \delta$ (mind that $\alpha \delta = \epsilon_v - 1$ here), \cite{zaslavsky2021first} reported consistency with $\alpha \delta = 1.3$, while we report $\epsilon_v-1 = \SI[separate-uncertainty]{1.04 \pm 0.20}{}$. 

As for comparison of present results with $\beta$-meteoroid flux near $\SI{1}{AU}$, \cite{wehry1999identification} reported the flux of $\beta$-meteoroids in ecliptic plane detected by Ulysses between $1.0 - 1.6 \si{AU}$ to be $\SI[separate-uncertainty]{1.5(3)e-4}{m^{-2} s^{-1}}$. \cite{zaslavsky2012interplanetary} reported flux of $\beta$-meteoroids at $\SI{1}{AU}$ of size $100 - 300 \si{nm}$ on STEREO/Waves in the range of $1 - 6 \times \SI{e-5}{m^{-2} s^{-1}}$, a somewhat lower value than reported here. SolO detections are likely of $\SI{100}{nm}$ and larger dust, but upper limit is somewhat higher for SolO due to wider dynamic range ($3 - 150 \si{mV}$ for STEREO and $3 - 700 \si{mV}$ for SolO), which may account for some of the difference. \cite{malaspina2015revisiting} however reported the value for STEREO/Waves by about a factor of $2.5$ higher than \cite{zaslavsky2012interplanetary}, which is on its upper bound virtually identical with the value reported here. For the Wind/WAVES experiment, \cite{malaspina2014interplanetary} reported $\SI[separate-uncertainty]{2.7(14) e-5 }{m^{-2} s^{-1}}$ for the sum of $\beta$-meteoroids and interstellar dust of $0.1 - 11\si{\mu m}$ size. It remains open if and how much interstellar dust contributes to the measurements of SolO/RPW analyzed in the present work. Recently, \cite{szalay2021collisional} reported $4 - 8 \cdot \SI{e-5}{m^{-2} s^{-1}}$ for $\beta$-meteoroid flux at $\SI{1}{AU}$ measured with Parker Solar Probe. The upper bound of this estimate is similar to the value reported in the present work. 
 
\section{Conclusions} \label{ch:conclusion}

   \begin{enumerate}
      \item We presented the analysis of the velocity of hyperbolic dust grains between $\SI{0.5}{AU} - \SI{1}{AU}$ based on highest-quality available data on daily dust detections by SolO/RPW, including discussion of implications for the velocity in case of non-hyperbolic (be it other dust population or false detections) component to the counts. Velocities in the range $30 - 110 \si{km/s}$ are compatible with the data.
      
      \item We presented a Bayesian hierarchical model and demonstrated how it is used to infer physical parameters of hyperbolic dust population in the studied region. It is likely that $\SI[separate-uncertainty]{1.5 \pm 0.3}{h^{-1}}$ are in fact not caused by hyperbolic dust. Then observations are consistent with a mean radial velocity of the hyperbolic component $\SI[separate-uncertainty]{63 \pm 7}{km/s}$ between $\SI{0.5}{AU}$ and $\SI{1}{AU}$. Spatial dependence of the detection rate suggests substantial deceleration of the observed hyperbolic dust particles. If they are $\beta$-meteoroids, the value of $\beta$ is likely just above the liberation threshold, specifically $\beta \gtrsim 0.5$ under the assumption of circular orbits of parent bodies. Hence closer to their origin, they likely have velocities higher than the inferred $\SI[separate-uncertainty]{63 \pm 7}{km/s}$.
      
      \item As a result of our modelling, we provide estimates of hyperbolic dust flux at $\SI{1}{AU}$ of $\SI[separate-uncertainty]{1.1(2) e-4}{m^{-2}s^{-1}}$, a value compatible with other relevant measurements. 
      
   \end{enumerate}
   
\section{Outlook} \label{ch:outlook}

SolO will get significant inclination in 2025, which will require further generalization of the model, to account for dust distribution out of the ecliptic plane. Out of ecliptic parameters of hyperbolic dust will likely provide more information on in-ecliptic hyperbolic dust, such as its parent bodies' mean eccentricity. Due to independence on $\beta$, knowledge of azimuthal velocity would be a good indicator of $\beta$-meteoroids origin, but it is hard to infer as the azimuthal component is much smaller than the radial component. Out of ecliptic detections may help in this regard as well.

Near the solar minimum of 2020, a so-called de-focusing solar magnetic field configuration was present in the solar system \citep{mann2010interstellar}. This configuration favours dust depletion of the near-ecliptic region due to the Lorentz force acting on ISD particles entering the solar system. With solar cycle 25, focusing field configuration will return at some time before the solar minimum of 2031. It is possible that a significant ISD component will be observed in years following the solar maximum of 2025, which will, if observed, provide new opportunities for dust population discrimination and more comprehensive dust cloud description thanks to SolO/RPW data. 


\begin{acknowledgements}
      Author contributions: Concept: SK, AT, IM. Data analysis: SK, SHS, AK. Interpretation: SK, IM, AT, SHS, AK, AZ. Manuscript preparation: SK. 
      
      The code and the data used in present work are publicly available at \url{github.com/SamuelKo1607/solo_dust_2022}.
      
      This work made use of publicly available data provided by A. Kvammen: \url{github.com/AndreasKvammen/ML_dust_detection}.
      
      SK and AT are supported by the Tromsø Research Foundation under the grant 19-SG-AT.
      
      This work is supported by the Research Council of Norway (grant number 262941). 
      
      Authors sincerely appreciate the support of Solar Orbiter/RPW Investigation team.
      
      This work was made possible by R-INLA package, authors thank to R-INLA team, see \url{r-inla.org}. 
\end{acknowledgements}


\newpage

\bibliographystyle{aa}
\renewcommand{\bibname}{References}
\bibliography{refs}

\begin{thebibliography}{58}
\expandafter\ifx\csname natexlab\endcsname\relax\def\natexlab#1{#1}\fi

\bibitem[{Alexander \& Bohn(1968)}]{alexander1968zodiacal}
Alexander, W. \& Bohn, J. 1968, in COSPAR Plenary Meeting No.
  NSSDC-ID-66-049A-21-PM, North-Holland Publishing co.,

\bibitem[{Baguhl {et~al.}(1996)Baguhl, Gr{\"u}n, \& Landgraf}]{baguhl1996situ}
Baguhl, M., Gr{\"u}n, E., \& Landgraf, M. 1996, Space Science Reviews, 78, 165

\bibitem[{Collette {et~al.}(2014)Collette, Gr{\"u}n, Malaspina, \&
  Sternovsky}]{collette2014micrometeoroid}
Collette, A., Gr{\"u}n, E., Malaspina, D., \& Sternovsky, Z. 2014, Journal of
  Geophysical Research: Space Physics, 119, 6019

\bibitem[{Czechowski \& Mann(2021)}]{czechowski2021dynamics}
Czechowski, A. \& Mann, I. 2021, Astronomy \& Astrophysics, 652, A131

\bibitem[{Dietzel {et~al.}(1973)Dietzel, Eichhorn, Fechtig, Grun, Hoffmann, \&
  Kissel}]{dietzel1973heos}
Dietzel, H., Eichhorn, G., Fechtig, H., {et~al.} 1973, Journal of Physics E:
  Scientific Instruments, 6, 209

\bibitem[{Dietzel {et~al.}(1972)Dietzel, Neukum, \&
  Rauser}]{dietzel1972micrometeoroid}
Dietzel, H., Neukum, G., \& Rauser, P. 1972, Journal of Geophysical Research,
  77, 1375

\bibitem[{Dohnanyi(1970)}]{dohnanyi1970origin}
Dohnanyi, J. 1970, Journal of Geophysical Research, 75, 3468

\bibitem[{Dohnanyi(1972)}]{dohnanyi1972interplanetary}
Dohnanyi, J. 1972, Icarus, 17, 1

\bibitem[{Friichtenicht(1962)}]{friichtenicht1962two}
Friichtenicht, J. 1962, Review of Scientific Instruments, 33, 209

\bibitem[{Gasque {et~al.}(2022)Gasque, Bale, Bowen, de~Wit, Geotz, Malaspina,
  Pusack, \& Szalay}]{gasque2022magnetic}
Gasque, C., Bale, S., Bowen, T., {et~al.} 2022, {AGU} Fall Meeting 2021

\bibitem[{G{\'o}mez-Rubio(2020)}]{gomez2020bayesian}
G{\'o}mez-Rubio, V. 2020, Bayesian inference with INLA (CRC Press)

\bibitem[{Gr{\"u}n(1984)}]{grun1984impact}
Gr{\"u}n, E. 1984, The Giotto Spacecraft Impact-induced Plasma Environment, 39

\bibitem[{Gr{\"u}n {et~al.}(2007)Gr{\"u}n, Pawlinka, \& Srama}]{grun2007dust}
Gr{\"u}n, E., Pawlinka, S., \& Srama, R. 2007, Max-Planck-Institut f{\"u}r
  Kernphysik

\bibitem[{Gr{\"u}n {et~al.}(1985)Gr{\"u}n, Zook, Fechtig, \&
  Giese}]{grun1985collisional}
Gr{\"u}n, E., Zook, H.~A., Fechtig, H., \& Giese, R. 1985, Icarus, 62, 244

\bibitem[{Gurnett {et~al.}(1997)Gurnett, Ansher, Kurth, \&
  Granroth}]{gurnett1997micron}
Gurnett, D., Ansher, J., Kurth, W., \& Granroth, L. 1997, Geophysical research
  letters, 24, 3125

\bibitem[{Howard {et~al.}(2019)Howard, Vourlidas, Bothmer, Colaninno, DeForest,
  Gallagher, Hall, Hess, Higginson, Korendyke, {et~al.}}]{howard2019near}
Howard, R.~A., Vourlidas, A., Bothmer, V., {et~al.} 2019, Nature, 576, 232

\bibitem[{Kurth {et~al.}(2006)Kurth, Averkamp, Gurnett, \&
  Wang}]{kurth2006cassini}
Kurth, W., Averkamp, T., Gurnett, D., \& Wang, Z. 2006, Planetary and Space
  Science, 54, 988

\bibitem[{Kvammen {et~al.}(2022)Kvammen, Wickstr{\o}m, Kociscak, Vaverka,
  Nouzak, Zaslavsky, Rackovic, Gjelsvik, Pisa, Soucek, \&
  Mann}]{kvammen2022convolutional}
Kvammen, A., Wickstr{\o}m, K., Kociscak, S., {et~al.} 2022, EGUsphere
  [preprint], 2022

\bibitem[{Leinert {et~al.}(1981)Leinert, Richter, Pitz, \&
  Planck}]{leinert1981zodiacal}
Leinert, C., Richter, I., Pitz, E., \& Planck, B. 1981, Astronomy and
  Astrophysics, 103, 177

\bibitem[{Maksimovic {et~al.}(2020)Maksimovic, Bale, Chust, Khotyaintsev,
  Krasnoselskikh, Kretzschmar, Plettemeier, Rucker, Sou{\v{c}}ek, Steller,
  {et~al.}}]{maksimovic2020solar}
Maksimovic, M., Bale, S., Chust, T., {et~al.} 2020, Astronomy \& Astrophysics,
  642, A12

\bibitem[{Malaspina {et~al.}(2014)Malaspina, Hor{\'a}nyi, Zaslavsky, Goetz,
  Wilson~III, \& Kersten}]{malaspina2014interplanetary}
Malaspina, D., Hor{\'a}nyi, M., Zaslavsky, A., {et~al.} 2014, Geophysical
  Research Letters, 41, 266

\bibitem[{Malaspina {et~al.}(2015)Malaspina, O'Brien, Thayer, Sternovsky, \&
  Collette}]{malaspina2015revisiting}
Malaspina, D.~M., O'Brien, L.~E., Thayer, F., Sternovsky, Z., \& Collette, A.
  2015, Journal of Geophysical Research: Space Physics, 120, 6085

\bibitem[{Malaspina {et~al.}(2022)Malaspina, Stenborg, Mehoke, Al-Ghazwi, Shen,
  Hsu, Iyer, Bale, \& de~Wit}]{malaspina2022clouds}
Malaspina, D.~M., Stenborg, G., Mehoke, D., {et~al.} 2022, The Astrophysical
  Journal, 925, 27

\bibitem[{Malaspina {et~al.}(2020)Malaspina, Szalay, Pokorn{\`y}, Page, Bale,
  Bonnell, de~Wit, Goetz, Goodrich, Harvey, {et~al.}}]{malaspina2020situ}
Malaspina, D.~M., Szalay, J.~R., Pokorn{\`y}, P., {et~al.} 2020, The
  Astrophysical Journal, 892, 115

\bibitem[{Malhotra \& Wang(2016)}]{malhotra2009eccentricity}
Malhotra, R. \& Wang, X. 2016, Monthly Notices of the Royal Astronomical
  Society, 465, 4381

\bibitem[{Mann(2010)}]{mann2010interstellar}
Mann, I. 2010, Annual Review of Astronomy and Astrophysics, 48, 173

\bibitem[{Mann \& Czechowski(2005)}]{mann2005dust}
Mann, I. \& Czechowski, A. 2005, The Astrophysical Journal, 621, L73

\bibitem[{Mann \& Czechowski(2021)}]{mann2021dust}
Mann, I. \& Czechowski, A. 2021, Astronomy \& Astrophysics, 650, A29

\bibitem[{Mann {et~al.}(2014)Mann, Meyer-Vernet, \& Czechowski}]{mann2014dust}
Mann, I., Meyer-Vernet, N., \& Czechowski, A. 2014, Physics reports, 536, 1

\bibitem[{Mann {et~al.}(2019)Mann, Nouzak, Vaverka, Antonsen, Fredriksen,
  Issautier, Malaspina, Meyer-Vernet, Pavlu, Sternovsky, Stude, Ye, \&
  Zaslavsky}]{mann2019dust}
Mann, I., Nouzak, L., Vaverka, J., {et~al.} 2019, Annales Geophysicae, 37, 1121

\bibitem[{Marshall \& Spiegelhalter(2003)}]{marshall2003approximate}
Marshall, E. \& Spiegelhalter, D. 2003, Statistics in medicine, 22, 1649

\bibitem[{Martins {et~al.}(2013)Martins, Simpson, Lindgren, \&
  Rue}]{martins2013bayesian}
Martins, T.~G., Simpson, D., Lindgren, F., \& Rue, H. 2013, Computational
  Statistics \& Data Analysis, 67, 68

\bibitem[{Meyer-Vernet {et~al.}(1986)Meyer-Vernet, Aubier, \&
  Pedersen}]{meyer1986voyager}
Meyer-Vernet, N., Aubier, M., \& Pedersen, B. 1986, Geophysical research
  letters, 13, 617

\bibitem[{Meyer-Vernet {et~al.}(2017)Meyer-Vernet, Moncuquet, Issautier, \&
  Schippers}]{meyer2017frequency}
Meyer-Vernet, N., Moncuquet, M., Issautier, K., \& Schippers, P. 2017, Journal
  of Geophysical Research: Space Physics, 122, 8

\bibitem[{Mozer {et~al.}(2020)Mozer, Agapitov, Bale, Bonnell, Goetz, Goodrich,
  Gore, Harvey, Kellogg, Malaspina, {et~al.}}]{mozer2020time}
Mozer, F., Agapitov, O., Bale, S., {et~al.} 2020, The Astrophysical Journal
  Supplement Series, 246, 50

\bibitem[{Nadaraya(1964)}]{nadaraya1964estimating}
Nadaraya, E.~A. 1964, Theory of Probability \& Its Applications, 9, 141

\bibitem[{Nouzk {et~al.}(2021)Nouzk, James, Nemecek, Safrankova, Pavlu,
  Novakova, Vaverka, \& Sternovsky}]{nouzak2021detection}
Nouzk, L., James, D., Nemecek, Z., {et~al.} 2021, The Astrophysical Journal,
  909, 132

\bibitem[{Pettit(1990)}]{pettit1990conditional}
Pettit, L. 1990, Journal of the Royal Statistical Society: Series B
  (Statistical Methodology), 52, 175

\bibitem[{Rackovic~Babic {et~al.}(2022)Rackovic~Babic, Zaslavsky, Issautier,
  Meyer-Vernet, \& Onic}]{babic2022analytical}
Rackovic~Babic, K., Zaslavsky, A., Issautier, K., Meyer-Vernet, N., \& Onic, D.
  2022, Astronomy \& Astrophysics, 659, A15

\bibitem[{Rue {et~al.}(2009)Rue, Martino, \& Chopin}]{rue2009approximate}
Rue, H., Martino, S., \& Chopin, N. 2009, Journal of the Royal Statistical
  Society: Series B (Statistical Methodology), 71, 319

\bibitem[{Rue {et~al.}(2017)Rue, Riebler, S{\o}rbye, Illian, Simpson, \&
  Lindgren}]{rue2017bayesian}
Rue, H., Riebler, A., S{\o}rbye, S.~H., {et~al.} 2017, Annual Review of
  Statistics and Its Application, 4, 395

\bibitem[{Shen {et~al.}(2021)Shen, Sternovsky, Garzelli, \&
  Malaspina}]{shen2021electrostatic}
Shen, M.~M., Sternovsky, Z., Garzelli, A., \& Malaspina, D.~M. 2021, Journal of
  Geophysical Research: Space Physics, 126, e2021JA029645

\bibitem[{Shu {et~al.}(2012)Shu, Collette, Drake, Gr{\"u}n, Hor{\'a}nyi, Kempf,
  Mocker, Munsat, Northway, Srama, {et~al.}}]{shu20123}
Shu, A., Collette, A., Drake, K., {et~al.} 2012, Review of Scientific
  Instruments, 83, 075108

\bibitem[{Solar Orbiter/RPW~Investigation(2022)}]{rwp_db}
Solar Orbiter/RPW~Investigation, L. 2022, Solar Orbiter / Radio and Plasma
  Waves Data, data retrieved from Observatoire de Paris, LESIA,
  \url{https://rpw.lesia.obspm.fr/roc/data/pub/solo/rpw/data/L2/}

\bibitem[{Srama {et~al.}(2004)Srama, Ahrens, Altobelli, Auer, Bradley, Burton,
  Dikarev, Economou, Fechtig, G{\"o}rlich, {et~al.}}]{srama2004cassini}
Srama, R., Ahrens, T.~J., Altobelli, N., {et~al.} 2004, The Cassini-Huygens
  Mission, 465

\bibitem[{Stenborg {et~al.}(2021)Stenborg, Howard, Hess, \&
  Gallagher}]{stenborg2021psp}
Stenborg, G., Howard, R., Hess, P., \& Gallagher, B. 2021, Astronomy \&
  Astrophysics, 650, A28

\bibitem[{Szalay {et~al.}(2021)Szalay, Pokorn{\`y}, Malaspina, Pusack, Bale,
  Battams, Gasque, Goetz, Kr{\"u}ger, McComas,
  {et~al.}}]{szalay2021collisional}
Szalay, J., Pokorn{\`y}, P., Malaspina, D., {et~al.} 2021, The Planetary
  Science Journal, 2, 185

\bibitem[{Van~de Hulst(1947)}]{van1947zodiacal}
Van~de Hulst, H. 1947, Astrophysical Journal, 105

\bibitem[{Vaverka {et~al.}(2018)Vaverka, Nakamura, Kero, Mann, De~Spiegeleer,
  Hamrin, Norberg, Lindqvist, \& Pellinen-Wannberg}]{vaverka2018comparison}
Vaverka, J., Nakamura, T., Kero, J., {et~al.} 2018, Journal of Geophysical
  Research: Space Physics, 123, 6119

\bibitem[{Vaverka {et~al.}(2017)Vaverka, Pellinen-Wannberg, Kero, Mann,
  De~Spiegeleer, Hamrin, Norberg, \& Pitk{\"a}nen}]{vaverka2017potential}
Vaverka, J., Pellinen-Wannberg, A., Kero, J., {et~al.} 2017, IEEE Transactions
  on Plasma Science, 45, 2048

\bibitem[{Wang {et~al.}(2006)Wang, Gurnett, Averkamp, Persoon, \&
  Kurth}]{wang2006characteristics}
Wang, Z., Gurnett, D., Averkamp, T., Persoon, A., \& Kurth, W. 2006, Planetary
  and Space Science, 54, 957

\bibitem[{Watson(1964)}]{watson1964smooth}
Watson, G.~S. 1964, Sankhy{\=a}: The Indian Journal of Statistics, Series A,
  359

\bibitem[{Wehry \& Mann(1999)}]{wehry1999identification}
Wehry, A. \& Mann, I. 1999, Astronomy and Astrophysics, 341, 296

\bibitem[{Whipple(1967)}]{whipple196756}
Whipple, F.~L. 1967, The Zodiacal Light and the Interplanetary Medium, 409

\bibitem[{Zaslavsky(2015)}]{zaslavsky2015floating}
Zaslavsky, A. 2015, Journal of Geophysical Research: Space Physics, 120, 855

\bibitem[{Zaslavsky {et~al.}(2021)Zaslavsky, Mann, Soucek, Czechowski,
  P{\'\i}{\v{s}}a, Vaverka, Meyer-Vernet, Maksimovic, Lorf{\`e}vre, Issautier,
  {et~al.}}]{zaslavsky2021first}
Zaslavsky, A., Mann, I., Soucek, J., {et~al.} 2021, Astronomy \& Astrophysics,
  656, A30

\bibitem[{Zaslavsky {et~al.}(2012)Zaslavsky, Meyer-Vernet, Mann, Czechowski,
  Issautier, Le~Chat, Pantellini, Goetz, Maksimovic, Bale,
  {et~al.}}]{zaslavsky2012interplanetary}
Zaslavsky, A., Meyer-Vernet, N., Mann, I., {et~al.} 2012, Journal of
  Geophysical Research: Space Physics, 117

\bibitem[{Zook \& Berg(1975)}]{zook1975source}
Zook, H.~A. \& Berg, O.~E. 1975, Planetary and Space Science, 23, 183

\end{thebibliography}


\begin{appendix} 

\section{Single-particle velocity and spatial density} \label{app:single_particle_valocities}

For the purpose of figs. \ref{fig:plt:beta_azimuthal_velocity} and $\ref{fig:plt:beta_rad_velocity}$, dust grains are assumed to move within the ecliptics, liberated from initially circular orbit and with their motion governed by the gravity and solar radiation pressure only, therefore

\vspace{-10pt} 

\begin{align}
    \begin{split} 
    |v| &= \sqrt{v_0^2 + 2GM(1-\beta)\left( \frac{1}{r} - \frac{1}{r_0}\right)},
    \end{split} \\
    \begin{split}
    v_{tan} = v_0 \frac{r_0}{r},
    \end{split} \\
    \begin{split}
    v_{rad} = \sqrt{v^2 - v_{tan}^2},
    \end{split}
\end{align}
where $v_0$ is the initial (purely radial) velocity and $r_0$ is the initial heliocentric distance (radius of the circular orbit). Furthermore, given a radial velocity profile of a radially escaping dust grain $v_{rad}(r)$, dust spatial density $\rho$ at a heliocentric distance $r$ is

\vspace{-10pt} 

\begin{align}
    \begin{split} 
    \rho (r) &= \rho (r_0) \left( \frac{r_0}{r} \right)^2 \frac{v_{rad}(r_0)}{v_{rad}(r)},
    \end{split}
\end{align}
where $r_0$ is a reference heliocentric distance.

\section{Dust radial velocity estimation} \label{app:velocity}

Suppose the detection rate is proportional to $v_{relative}^q$:
 
\vspace{-10pt} 
 
\begin{align}
    \begin{split} 
    R &= R_0 \cdot v_{relative}^q = R_0 \left( \vec{v_{dust}} - \vec{v_{sc}} \right)^q 
    \end{split} \\
    \begin{split}
    &= R_0 \left[ \sqrt{ ( v_{dust;rad} - v_{sc;rad})^2 + v_{sc;azim}^2 } \right]^{q},
    \end{split}
\end{align}
where we assumed $v_{dust;azim} = 0$. Then at any given heliocentric distance $r$:

\vspace{-10pt}

\begin{align}
    \begin{split} 
    R_{in}^{2/q} &= R_0^{2/q} \left( (v_{dust;rad} + |v_{sc;rad}|)^2 + v_{sc;azim}^2 \right),
    \end{split} \\
    \begin{split}
    R_{out}^{2/q} &= R_0^{2/q} \left( (v_{dust;rad} - |v_{sc;rad}|)^2 + v_{sc;azim}^2 \right),
    \end{split}
\end{align}
and therefore:

\vspace{-10pt}

\begin{equation}
    \frac{R_{in}^{2/q}}{R_{out}^{2/q}} = \frac{(v_{dust;rad} + |v_{sc;rad}|)^2 + v_{sc;azim}^2}{(v_{dust;rad} - |v_{sc;rad}|)^2 + v_{sc;azim}^2},
\end{equation}
from which:

\vspace{-10pt}

\begin{align}
    \begin{split} 
    0 &= v_{dust;rad}^2 \cdot \left( R_{in}^{2/q} - R_{out}^{2/q} \right)
    \end{split}\\
    \begin{split} 
    &+ v_{dust;rad} \cdot \left( -2v_{sc;rad} \left( R_{in}^{2/q} + R_{out}^{2/q} \right) \right)
    \end{split}\\
    \begin{split} 
    &+ \left( R_{in}^{2/q} - R_{out}^{2/q} \right) \cdot \left( v_{sc;rad}^2 + v_{sc;azim}^2 \right),
    \end{split}\\
\end{align}
which leads to a quadratic root:

\vspace{-10pt}

\begin{align}
    \begin{split} 
    v_{dust;rad} &= \frac{2 | v_{sc;rad} | \left( R_{in}^{2/q} + R_{out}^{2/q} \right) \pm \sqrt{D} }{2\left( R_{in}^{2/q} - R_{out}^{2/q} \right)}, \label{eq:ap1_solution}
    \end{split}\\
    \begin{split} 
     D &= 4v_{sc;rad}^{2} \left( R_{in}^{2/q} + R_{out}^{2/q} \right)^2 - 4v_{sc}^{2} \left( R_{in}^{2/q} - R_{out}^{2/q} \right)^2,
    \end{split}
\end{align}
where $(+)$ in eq. \eqref{eq:ap1_solution} leads to positive velocity $v_{sc;rad}$. It is easy to see that in the special case of $q=1; v_{sc;azim}=0$:

\vspace{-10pt}

\begin{align}
    \begin{split} 
     D &= 4v_{sc;rad}^{2} \left[ \left( R_{in}^{2} + R_{out}^{2} \right)^2 - \left( R_{in}^{2} - R_{out}^{2} \right)^2 \right],
    \end{split}\\
    \begin{split} 
     &= 16v_{sc;rad}^{2} R_{in}^{2} R_{out}^{2},
    \end{split}
\end{align}
and by extension:

\vspace{-10pt}

\begin{align}
    \begin{split} 
    v_{dust;rad} &= \frac{|v_{sc;rad}| \left[ \left( R_{in}^{2} + R_{out}^{2} \right) \pm 2 R_{in} R_{out} \right] }{\left( R_{in}^{2} - R_{out}^{2} \right)},
    \end{split}
\end{align}
which is 

\vspace{-10pt}

\begin{align}
    \begin{split} 
    v_{dust;rad} &= \frac{|v_{sc;rad}| \left( R_{in} + R_{out} \right)  }{\left( R_{in} - R_{out} \right)}
    \end{split}
\end{align}
for $(+)$ in numerator.

\newpage

\section{Velocity inference --- full velocity profiles} \label{app:full_velocity_profiles}

\begin{figure}[H]
\centering
\includegraphics{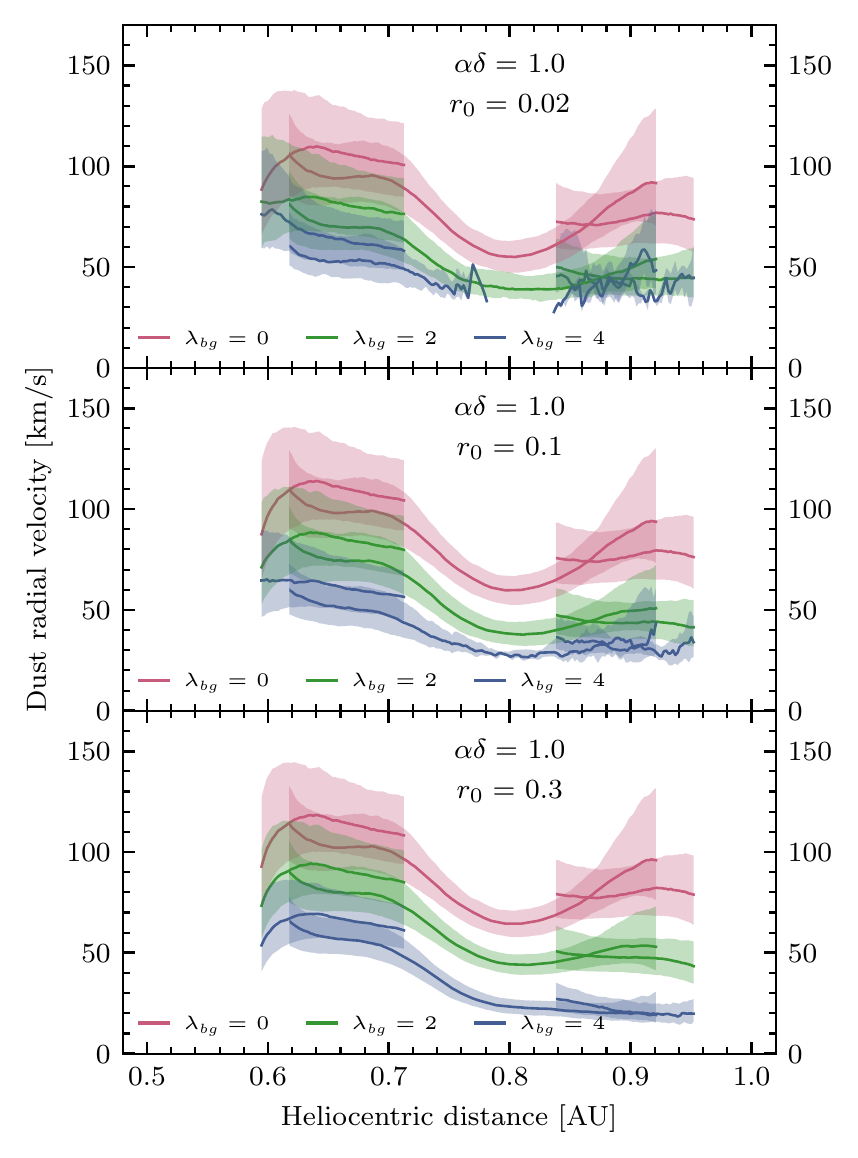}
\caption{Velocity estimated from TDS/TSWF-E/CNN under the assumption of $\alpha \delta = 1.0$. Panels correspond to different initial heliocentric distances. Colors correspond to different assumptions of background detection rate.}
\label{fig:plt:velocity_profile_bg0}
\end{figure}

\begin{figure}[H]
\centering
\includegraphics{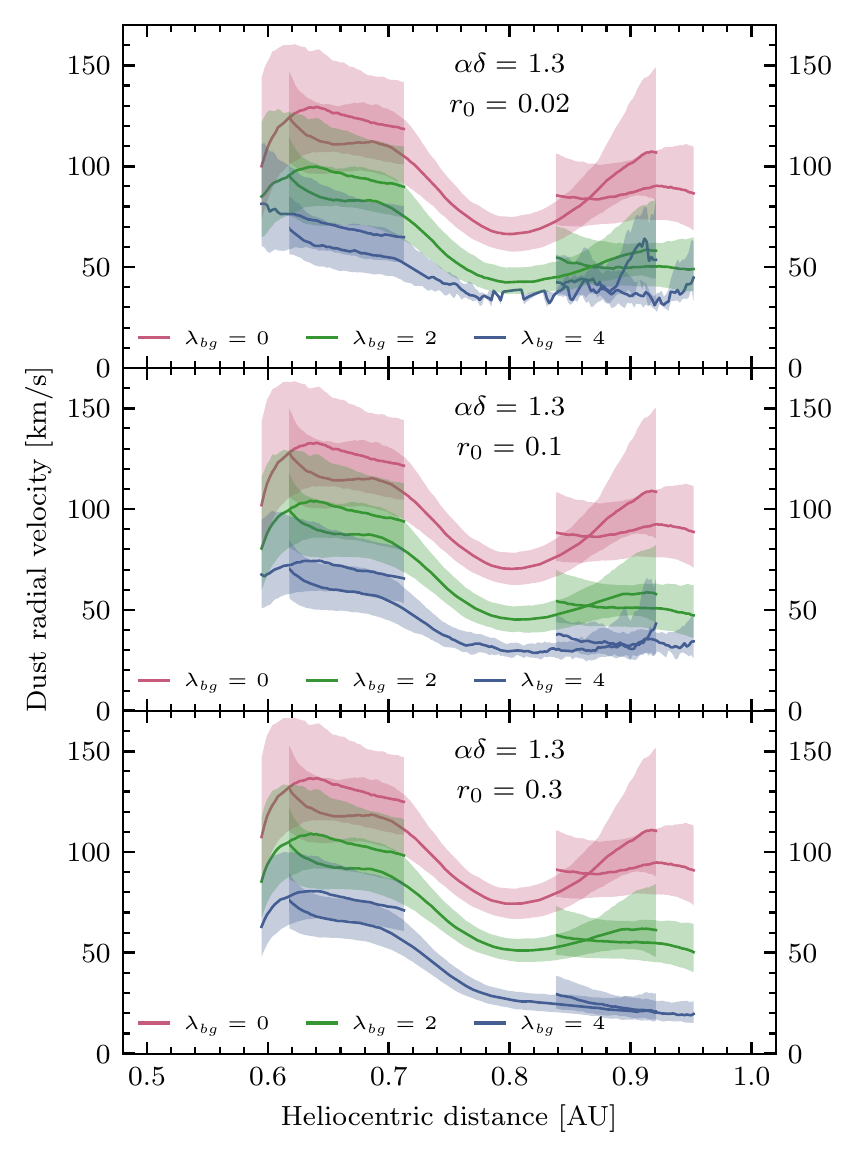}
\caption{Velocity estimated from TDS/TSWF-E/CNN under the assumption of $\alpha \delta = 1.3$. Panels correspond to different initial heliocentric distances. Colors correspond to different assumptions of background detection rate.}
\label{fig:plt:velocity_profile_bg2}
\end{figure}

\begin{figure}[H]
\centering
\includegraphics{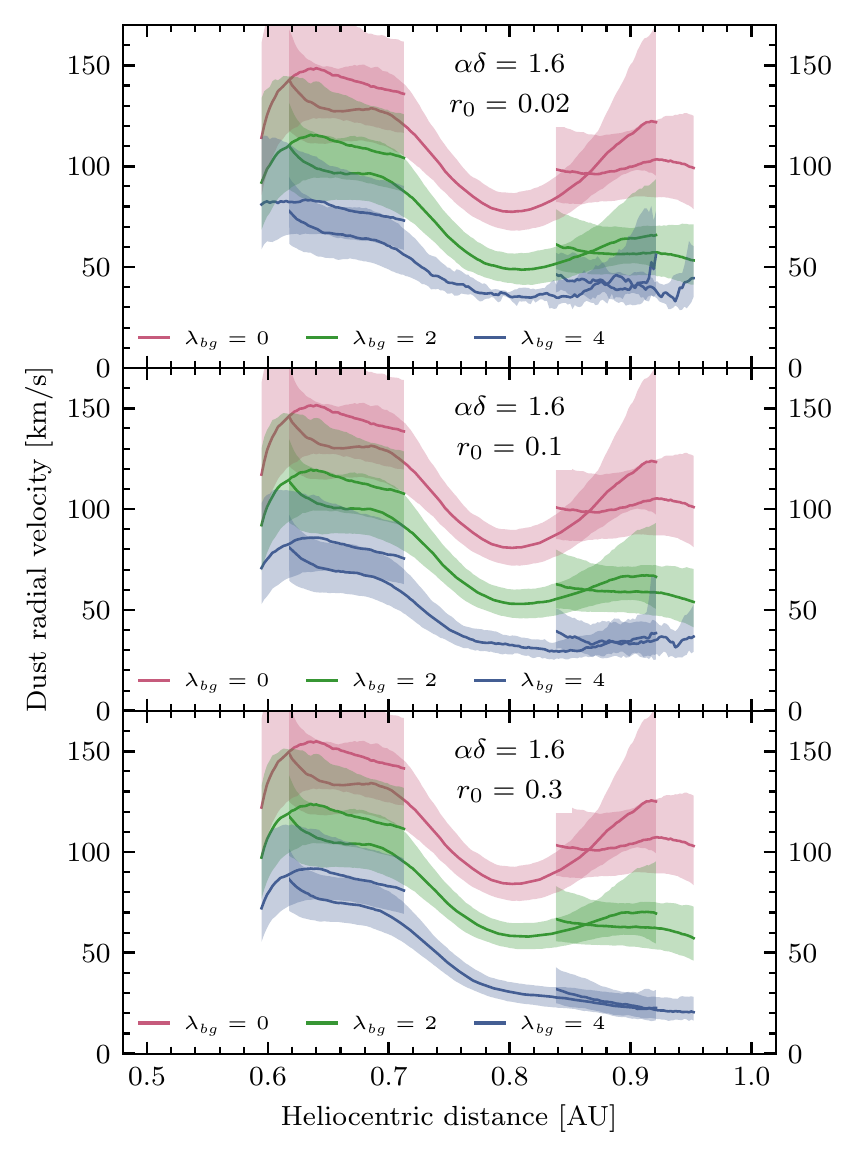}
\caption{Velocity estimated from TDS/TSWF-E/CNN under the assumption of $\alpha \delta = 1.6$. Panels correspond to different initial heliocentric distances. Colors correspond to different assumptions of background detection rate.}
\label{fig:plt:velocity_profile_bg4}
\end{figure}

The velocity profiles inferred in section \ref{ch:velocity_nonparametric} are shown in Figs. \ref{fig:plt:velocity_profile_bg0} to \ref{fig:plt:velocity_profile_bg4} (compare to Figs. \ref{fig:plt:velocity_profile_only_background} and \ref{fig:plt:velocity_other_parameters}). Note that the missing solutions (jittery line) for heliocentric distance $> \SI{0.7}{AU}$ and $\lambda_{bg}=4$ cause incomplete data shown in Fig. \ref{fig:plt:velocity_other_parameters}. These solutions only exist for some combinations of the free parameters, in particular for $\lambda_{bg} = 4$. 

\newpage
\section{Interpretation of the parameter $\epsilon_r$} \label{app:epsilon_r}

Intuitive explanation of the parameter $\nu_r$ as the mean dust velocity and of the factor $r^{\epsilon_r}$ as the spatial density can be clarified, assuming $\nu_r \propto r^\xi$. With the model \eqref{eq:model_rate}, the non-constant component $\tilde{R}$ of the rate $R$ is proportional to
\vspace{-10pt}
\begin{align}
    \begin{split} 
    \tilde{R} &\propto r^{\epsilon_r} \cdot v_{impact}^{\epsilon_v},
    \end{split}
\end{align}
where $\epsilon_r = -2$ in case of no acceleration of the dust. $\epsilon_v$ is explained as $\epsilon_v = 1 + \alpha \delta$, therefore
\vspace{-10pt}
\begin{align}
    \begin{split} 
    \tilde{R} &\propto r^{-2} \cdot v_{impact}^{1} \cdot v_{impact}^{\alpha \delta}.
    \end{split}
\end{align}
The factor of $v_{impact}^1$ actually comes from the proportionality
\vspace{-10pt}
\begin{align}
    \begin{split} 
    \tilde{R} &\propto \frac{v_{impact}}{\nu_r},
    \end{split}
\end{align}
if we assume only radial motion for simplicity, compare with eq. \eqref{eq:z2_model}. It is also apparent from the following: Assuming stationary spacecraft ($v_{impact} = \nu_r$), the detection rate (in $\si{s^{-1}}$) is a product of the flux $F(r)$ (in $\si{s^{-1}}$) and the detection area $S$ (in $\si{m^{2}}$), independently of $\nu_r$. We therefore have for non-accelerating dust: 
\vspace{-10pt}
\begin{align}
    \begin{split} 
    \tilde{R} &\propto r^{-2} \cdot \frac{v_{impact}}{\nu_r} \cdot v_{impact}^{\alpha \delta},
    \end{split}
\end{align}
and finally assuming $\nu_r \propto r^\xi$ we get 
\vspace{-10pt}
\begin{align}
    \begin{split} 
    \tilde{R} &\propto r^{-2} \cdot r^{-\xi} \cdot v_{impact} \cdot v_{impact}^{\alpha \delta} = r^{-2-\xi} \cdot v_{impact}^{\epsilon_v},
    \end{split}
    \label{eq:epsilon_r_explanation}
\end{align}
therefore 
\vspace{-10pt}
\begin{align}
    \begin{split} 
    \epsilon_r &= -2-\xi.
    \end{split}
\end{align}

There is a dichotomy in eq. \eqref{eq:epsilon_r_explanation} in that the assumption of $v_{dust} \propto r^\xi$ was used to expand the $v_{dust}$ but not the $v_{impact}$. This is one way of interpreting the approximation described in section \ref{ch:model_formulation}, we assume a non-constant dust velocity in the factor for spatial dust density, but a constant radial dust velocity in the expression for $v_{impact}$, see eq. \eqref{eq:model_relative_velocity}. This is done because of a clear relation of $\epsilon_r \lessgtr -2$ to acceleration and deceleration of the dust, which in our case ($\epsilon_r \approx -1.6 \implies \xi \approx -0.4$) reveals that the dust is decidedly decelerating.

\newpage

\section{Spatial density profiles} \label{app:distance_exponents_different_initial}

\begin{figure}[H]
\centering
\includegraphics{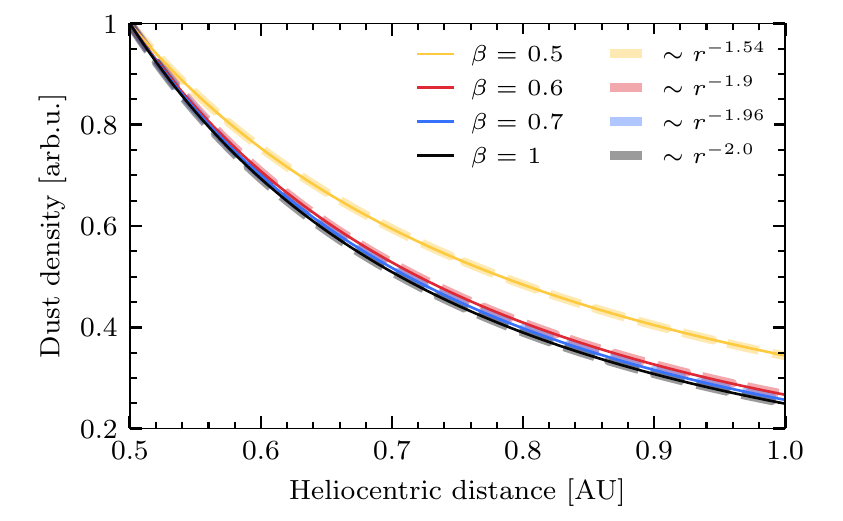}
\caption{Modelled dust spatial densities for different $\beta$ values assuming a circular initial orbit of $\SI{0.05}{AU}$. Solid lines show the spatial density and are normalized to the density at $\SI{0.5}{AU}$. Dashed lines are approximations to the solid lines, assuming power dependence on $r$.}
\label{fig:plt:distance_exponents_005}
\end{figure}

\begin{figure}[H]
\centering
\includegraphics{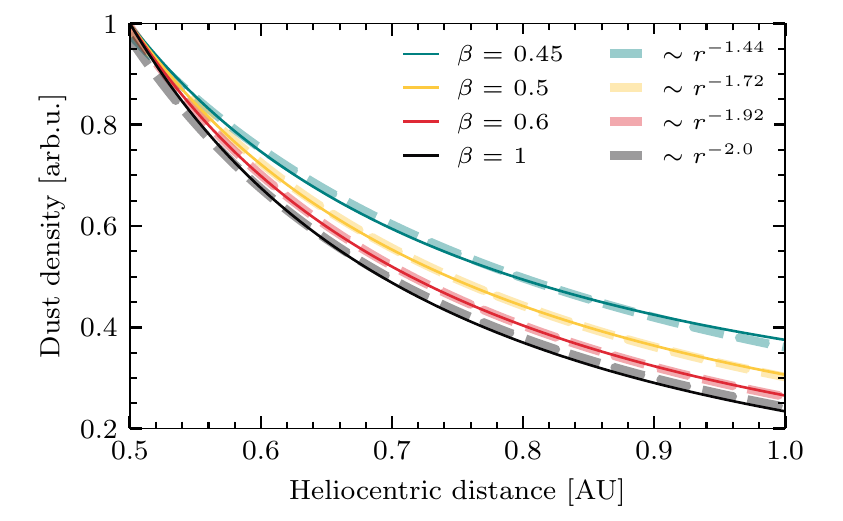}
\caption{Modelled dust spatial densities for different $\beta$ values assuming a circular initial orbit of $\SI{0.2}{AU}$. Solid lines show the spatial density and are normalized to the density at $\SI{0.5}{AU}$. Dashed lines are approximations to the solid lines, assuming power dependence on $r$.}
\label{fig:plt:distance_exponents_02}
\end{figure}

\begin{figure}[H]
\centering
\includegraphics{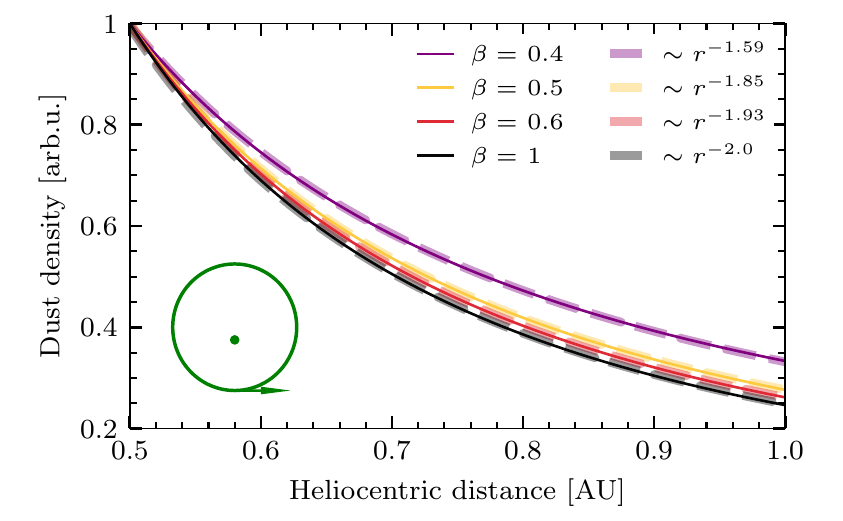}
\caption{Modelled dust spatial densities for different $\beta$ values assuming elliptical initial orbit with eccentricity of $e = 0.2$ and perihelion ejection. A perihelion of $\SI{0.1}{AU}$ is assumed. Solid lines show the spatial density and are normalized to the density at $\SI{0.5}{AU}$. Dashed lines are approximations to the solid lines, assuming power dependence on $r$.}
\label{fig:plt:distance_exponents_eccentric}
\end{figure}

Assumed initial orbital distance influences the relationship between $\beta$ values and spatial dust density profiles. In Fig. \ref{fig:plt:distance_exponents}, initial orbit of $\SI{0.1}{AU}$ is assumed. See plots \ref{fig:plt:distance_exponents_005} and \ref{fig:plt:distance_exponents_02} for similar plots with different assumptions of initial orbit. See Fig. \ref{fig:plt:distance_exponents_eccentric} for similar plot if eccentricity of $e = 0.2$ and perihelion ejection with perihelion of $r=\SI{0.1}{AU}$ is assumed. Although the estimates of the exponents vary, the general conclusion of lower $\beta$ implying lower exponent holds. Note that the profile gets significantly influenced when $\beta$ is close to threshold, and that it depends on the initial orbit and eccentricity. Therefore, $\beta < \SI{0.5}{}$ is shown where $\SI{0.5}{}$ is much higher than the liberation threshold. 

\newpage
\section{Model assessment} \label{app:model_evaluation}

There are several options for model evaluation implemented in R-INLA \citep[Chapter~2.4]{gomez2020bayesian}. We briefly describe two measures of choice. 

The conditional predictive ordinates (CPO, see \cite{pettit1990conditional}) for a given observation point gives the posterior probability of each observation when this observation is ommited in the model fitting. CPO is used to detect surprising observations or outliers. We examined the fit for failure flags for all the points, that would suggest a contradiction between the model and a data point. No failures were encountered.

\begin{figure}[H]
\centering
\includegraphics{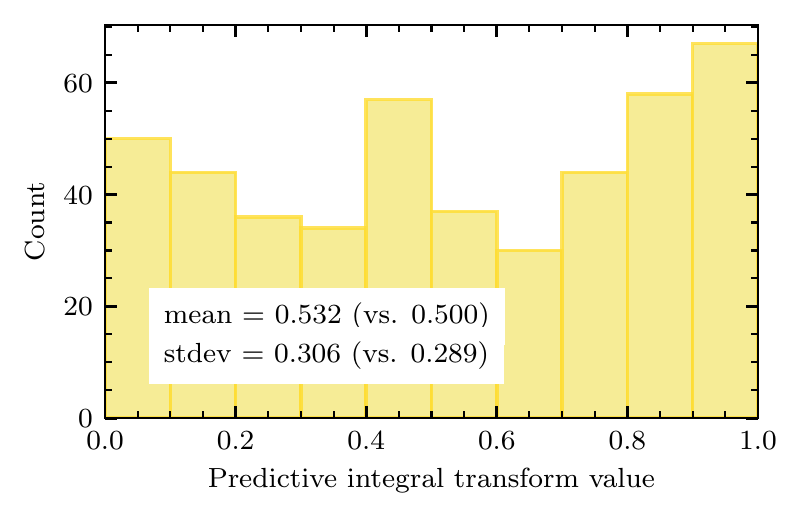}
\caption{Histogram of PIT values for the model described in section \ref{ch:posterior_distributions}. Mean and standard deviation are compared to the values for uniform distribution.}
\label{fig:plt:pit}
\end{figure}

The predictive integral transform (PIT, see \cite{marshall2003approximate}) measures the probability that a new observation will be lower than the observed value for each observation point individually. Histogram of the PIT values should therefore be similar to the uniform distribution between $0$ and $1$ when the model explains data well, see Fig. \ref{fig:plt:pit}.

\newpage
\section{Covariance plots of posteriors} \label{app:covariance_posteriors}

\begin{figure}[H]
\centering
\includegraphics{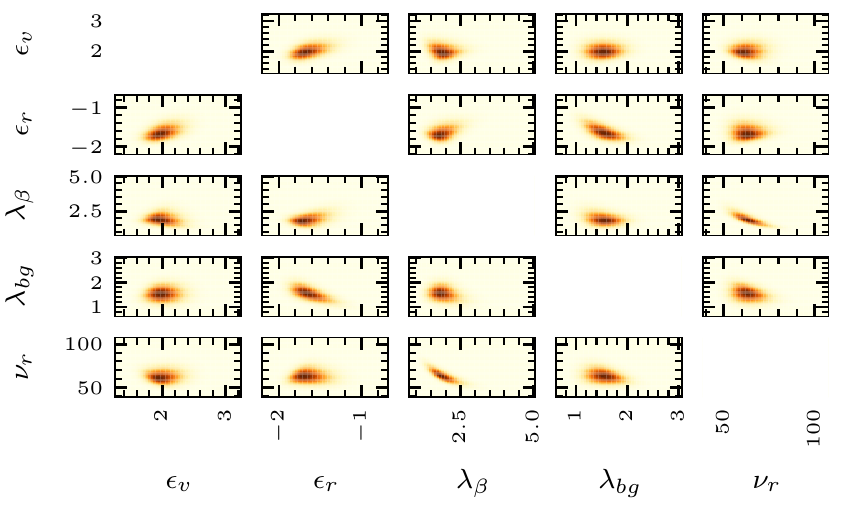}
\caption{Illustration of the covariance between all parameter pairs, constructed using sampling from the joint posterior distribution of all parameters.}
\label{fig:plt:all_covariances}
\end{figure}

As is shown in figure \ref{fig:plt:all_covariances}, basically all parameter pairs show substantial correlation. The pairs hold useful information, but are hardly surprising and are easy to interpret, having the model eq. \eqref{eq:model_rate} in mind. The correlation is unimportant in the case of $\lambda_\beta$, which has a role of a normalization constant. 

\begin{table}[H]
\centering
\begin{tabular}{r|c|c|c|c|c}
   & $\epsilon_v$  & $\epsilon_r$ & $\lambda_\beta$ & $\lambda_{bg}$ & $\nu_r$ \\ \hline
$\epsilon_v$        &      & 0.451 & -0.109& 0.068 & -0.015  \\ \hline
$\epsilon_r$        & 0.451 &      & 0.392 & -0.621& 0.027  \\ \hline
$\lambda_\beta$     & -0.109& 0.392 &      & -0.244& -0.799 \\ \hline
$\lambda_{bg}$      & 0.068 & -0.621& -0.244&      & -0.303 \\ \hline
$\nu_r$             & -0.015& -0.027& -0.799& -0.303& 
\end{tabular}
\caption{Covariance between all parameter pairs, constructed using sampling from the joint posterior distribution of all parameters.}
\label{tab:all_covariances}
\end{table}

\newpage
\section{Model fitting to original data} \label{app:tds_inla}

\begin{figure}[H]
\centering
\includegraphics{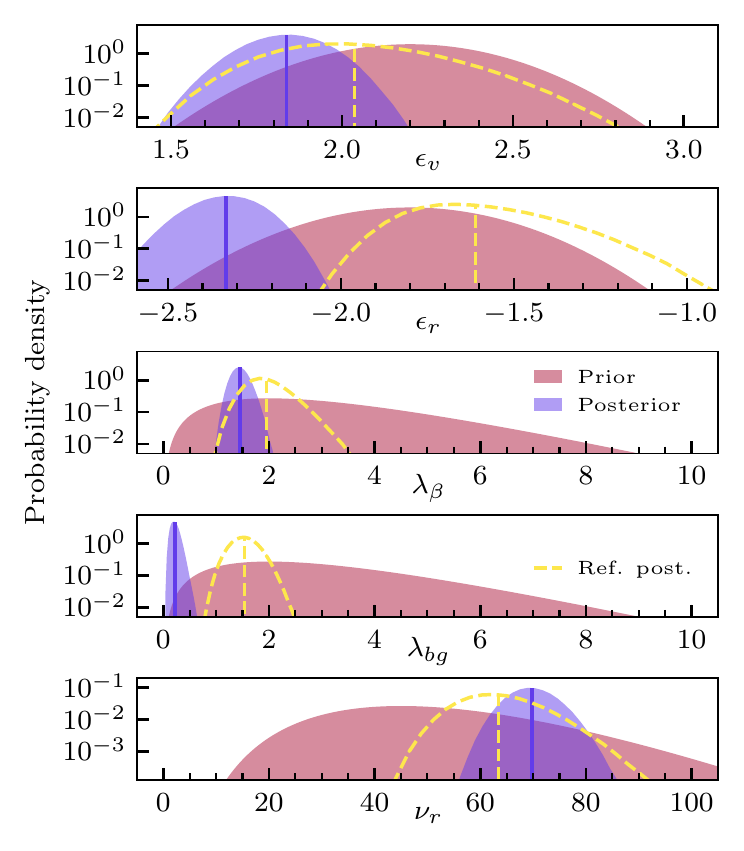}
\caption{Prior and posterior distributions of parameters, making use of original TDS (onboard processed) data. Prior distributions are described in text, sec. \ref{ch:inla}. Posterior distributions are described in Tab. \ref{tab:posteriors_tds}. Posteriors from Fig. \ref{fig:plt:priors_posteriors} are shown as reference in dashed lines for comparison.}
\label{fig:plt:priors_posteriors_tds}
\end{figure}

The procedure described in sec. \ref{ch:inla} was also applied to the original TDS data, meaning impacts identified onboard SolO, described by \cite{maksimovic2020solar}, which are different from TDS/TSWF-E/CNN data \citep{kvammen2022convolutional} used otherwise in sec. \ref{ch:inla}. We have every reason to believe that CNN-refined data used in sec. \ref{ch:inla} are more reliable, with fewer type 1 and type 2 errors, but the original data was used previously \citep{zaslavsky2021first} and the inspection of the result of the procedure is instructive nonetheless. The results are presented in Fig. \ref{fig:plt:priors_posteriors_tds}.

The most important and intuitive difference is that a much lower $\lambda_{bg}$ is inferred in this case (compare with Fig. \ref{fig:plt:priors_posteriors}). As described in \cite{kvammen2022convolutional}, the CNN procedure identifies substantially more impacts near-aphelion, which suggests more background dust, everything else being equal. Importantly, the inferred velocity $\nu_r$ doesn't change substantially, even though $\epsilon_v$ and especially $\epsilon_r$ do change consequentially. Importantly, $\epsilon_r < -2$ implies accelerating dust, which implies $\beta > 1$ and requires specific material and size of the grains, hence this is unlikely --- at least for $\beta$-meteoroids. Note that $\epsilon_r$ is effectively far from our prior expectations, providing poor fit to our prior knowledge. These results lend additional credence to the improvement of the CNN data.

\begin{table}[H]
\centering
\begin{tabular}{r|c|c}
   & $Mean$  & St. dev. \\ \hline
$\epsilon_v$ & 1.84  & 0.10 \\ \hline
$\epsilon_r$ & -2.33 & 0.09 \\ \hline
$\lambda_\beta$ & 1.45  & 0.16 \\ \hline
$\lambda_{bg}$ & 0.23  & 0.08 \\ \hline
$\nu_r$ & 69.8  & 4.06 
\end{tabular}
\caption{Summary statistics for the parameters, making use of original TDS (onboard processed) data. For visual representation see Fig. \ref{fig:plt:priors_posteriors_tds}.}
\label{tab:posteriors_tds}
\end{table}

\newpage
\section{Variation of priors and other parameters} \label{app:priors_variation}

\begin{figure}[H]
\centering
\includegraphics{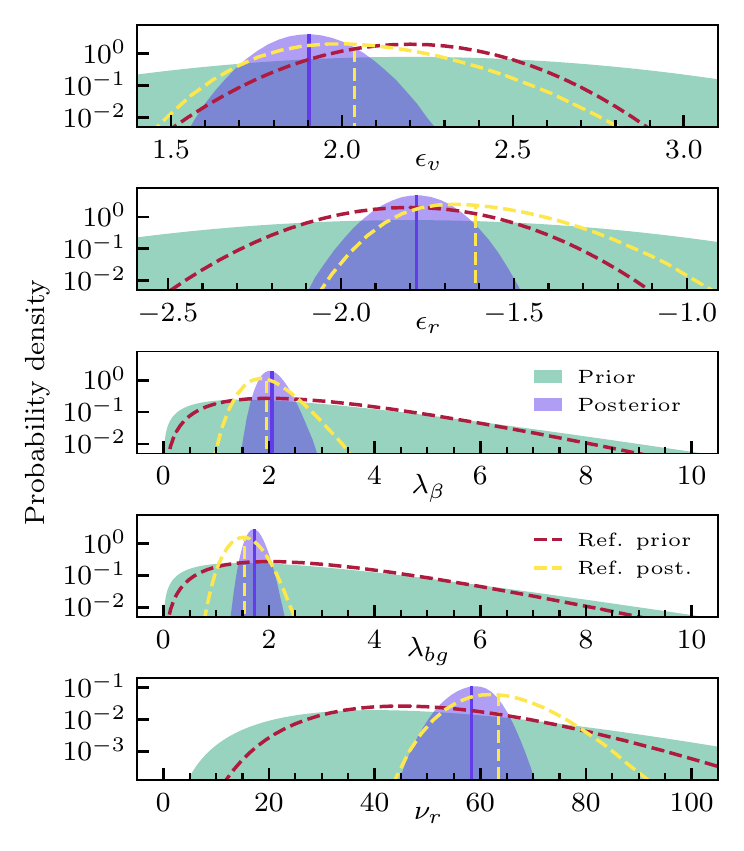}
\caption{Prior and posterior distributions of the parameters, the priors being substantially wider (less informative). Priors and posteriors from Fig. \ref{fig:plt:priors_posteriors} are shown in dashed lines for comparison.}
\label{fig:plt:priors_posteriors_vague}
\end{figure}

\begin{figure}[H]
\centering
\includegraphics{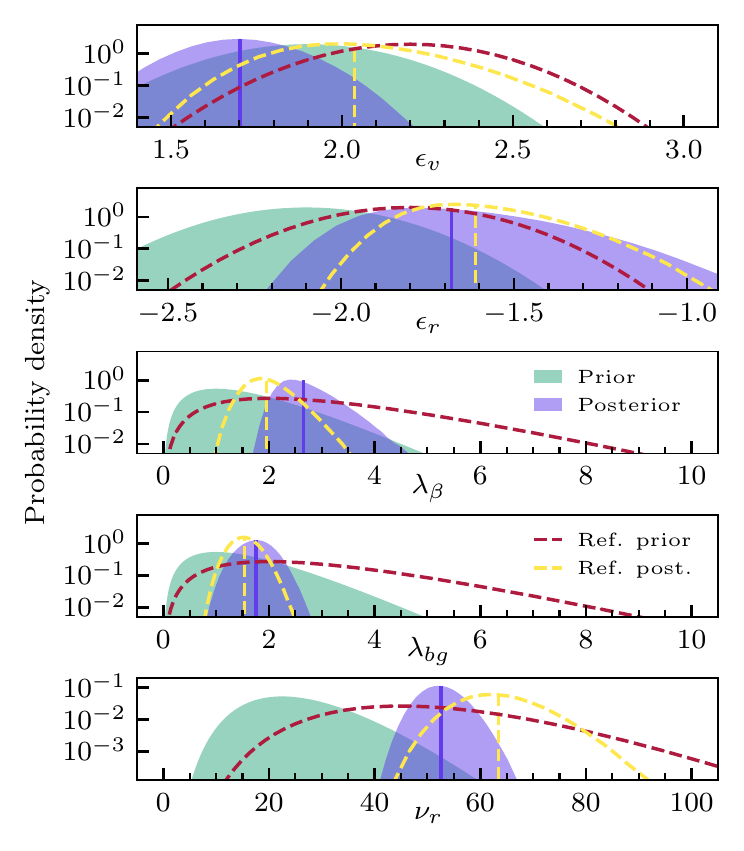}
\caption{Prior and posterior distributions of parameters, priors shifted towards lower values. Priors and posteriors from Fig. \ref{fig:plt:priors_posteriors} are shown in dashed lines for comparison.}
\label{fig:plt:priors_posteriors_to_left}
\end{figure}

\begin{figure}[H]
\centering
\includegraphics{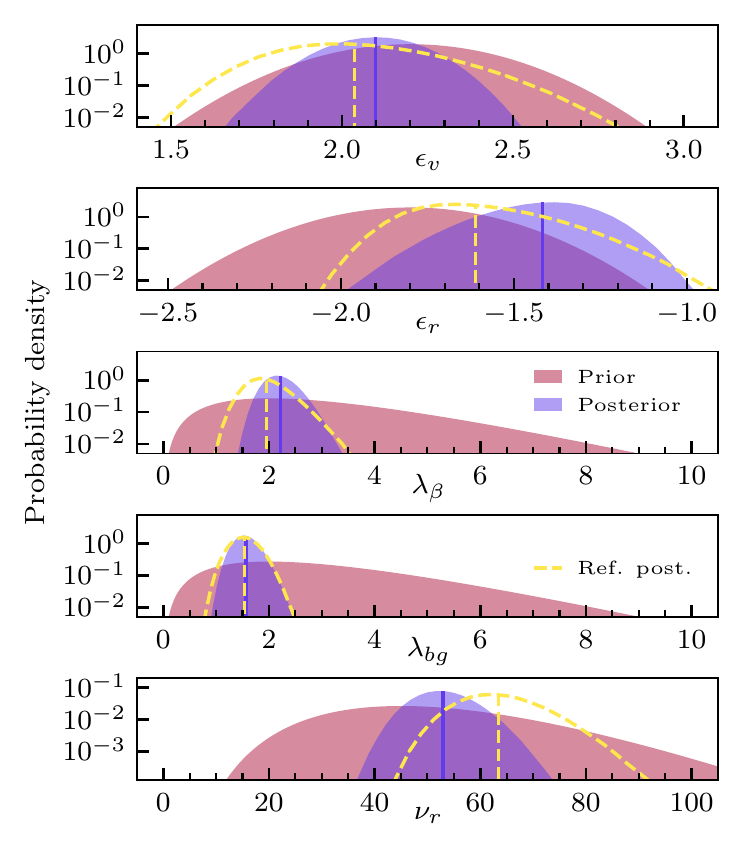}
\caption{Prior and posterior distributions of parameters, fixed parameter of azimuthal velocity is changed from $\SI{12}{km/s}$ at $\SI{0.75}{AU}$ to constant $\SI{0}{km/s}$. Posteriors from Fig. \ref{fig:plt:priors_posteriors} are shown in dashed lines for comparison.}
\label{fig:plt:priors_posteriors_half_vt}
\end{figure}

\begin{figure}[H]
\centering
\includegraphics{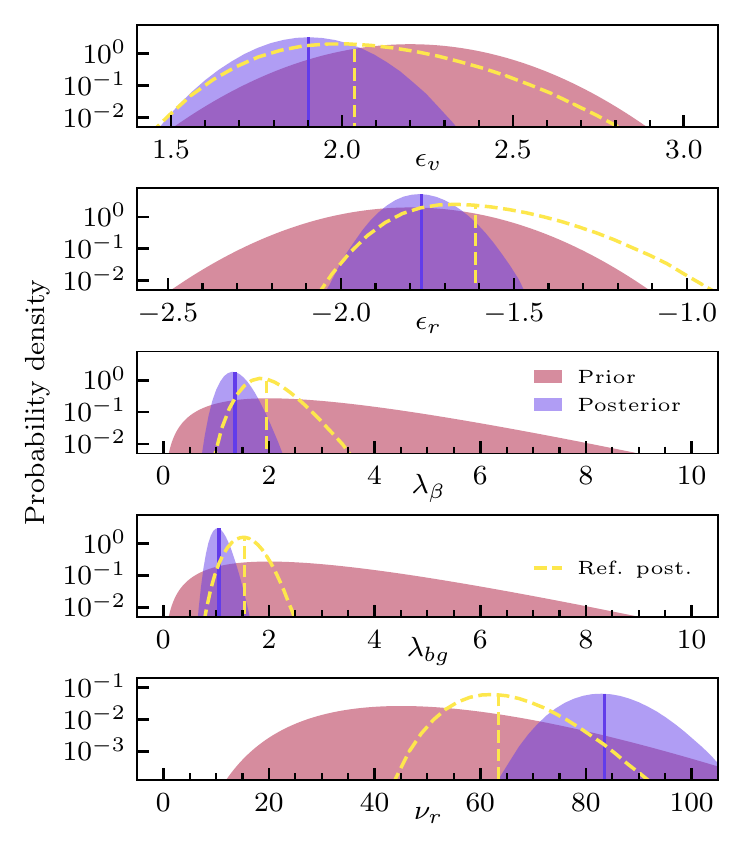}
\caption{Prior and posterior distributions of parameters, the fixed parameter of azimuthal velocity at $\SI{0.75}{AU}$ is changed from $\SI{12}{km/s}$ to $\SI{24}{km/s}$, which is a value higher by $\SI{100}{\%}$. Posteriors from Fig. \ref{fig:plt:priors_posteriors} are shown in dashed lines for comparison.}
\label{fig:plt:priors_posteriors_double_vt}
\end{figure}

\begin{figure}[H]
\centering
\includegraphics{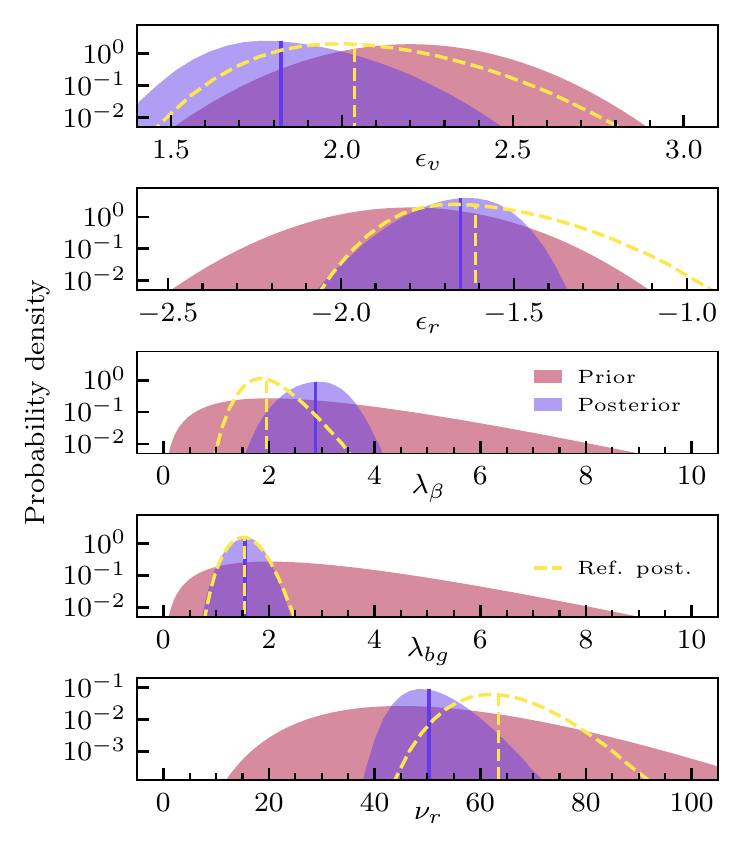}
\caption{Prior and posterior distributions of parameters, with starting point of $\epsilon_v = 3; \epsilon_r = -3; \lambda_\beta = 3; \lambda_{bg} = 3; \nu_r = 30$. Posteriors from Fig. \ref{fig:plt:priors_posteriors} are shown in dashed lines for comparison.}
\label{fig:plt:priors_posteriors_all_3_and_30}
\end{figure}

It is true that the priors themselves express the uncertainty in prior knowledge, but to demonstrate the robustness of the analysis, we here show perturbed priors and resulting posterior combinations (Figs. \ref{fig:plt:priors_posteriors_vague} and \ref{fig:plt:priors_posteriors_to_left}), to show that the result, although somewhat dependent on the prior selection, does not change dramatically if priors are chosen arbitrarily slightly different. Also the choice of the value of parameter $v_{a}$ (which is not a free parameter in our modelling, see eq. \eqref{eq:model_azimuthal_velocity}) is examined here (see Figs. \ref{fig:plt:priors_posteriors_half_vt} and \ref{fig:plt:priors_posteriors_double_vt}). Last but not least, we show the posteriors in case of change of the initialization of the iterative procedure to estimate the parameters (Fig. \ref{fig:plt:priors_posteriors_all_3_and_30}). We do not claim that any of these results is as trustworthy as the main result shown in Fig. \ref{fig:plt:priors_posteriors}, we had reasoning behind choosing the priors and parameters that we chose. Note that the mean of the marginal posteriors shown in Figs. \ref{fig:plt:priors_posteriors_vague} to \ref{fig:plt:priors_posteriors_all_3_and_30} lie within high credibility regions of posteriors shown in Fig. \ref{fig:plt:priors_posteriors} and vice versa, which supports the claim that the analysis presented here is robust. Observe that parameter values inferred with lower precision (wider posterior distributions) are more susceptible to change due to change in parameters, which is in line with expectations and with the meaning of precision here. Good choice of priors is still important for getting the highest quality estimate, but the result is not critically sensitive. 

\newpage
\section{Possible background profiles} \label{app:background}

\begin{figure}[H]
\centering
\includegraphics{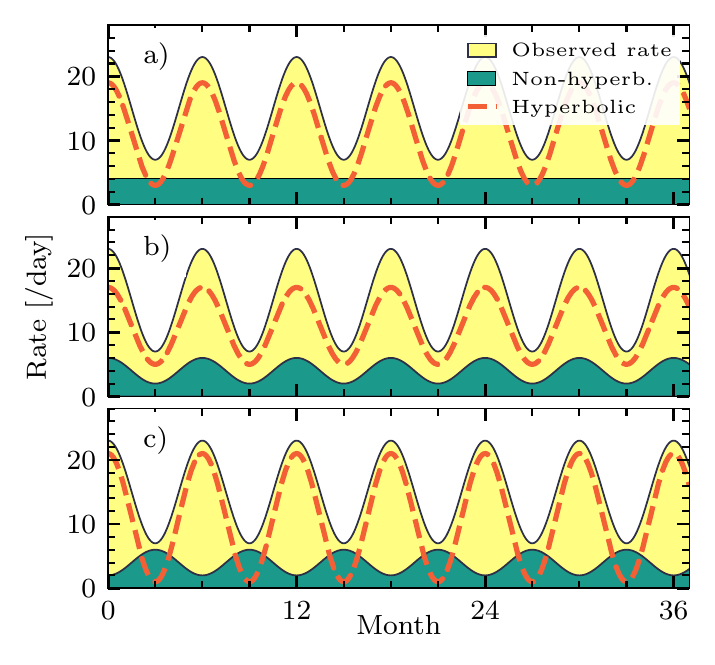}
\caption{Detection rate: selection of different combinations of hyperbolic and non-hyperbolic compounding to the same observed detection rate. In panel a) the background component is independent on heliocentric distance. In panel b) the non-hyperbolic component is negatively correlated with heliocentric distance. In panel c) the non-hyperbolic component is positively correlated with heliocentric distance.}
\label{fig:plt:background_profiles}
\end{figure}

In the present analysis, the non-hyperbolic component was assumed to be present and constant. See Fig. \ref{fig:plt:background_profiles} for examples of possible non-hyperbolic component profiles, as discussed in section \ref{ch:posterior_distribution}. In the plot, mean rate of non-hyperbolic component is the same in all three panels. Note that despite that, the dynamic range (that is ratios of maximum over minimum values) changes significantly as a result of the change in the temporal profile of the non-hyperbolic component. Significant deviation from the constant case would therefore likely change both the inferred prevalence of the non-hyperbolic component and the parameters of the hyperbolic grains.

\end{appendix}

\end{document}